\documentclass{article} 
\usepackage{iclr2027_conference,times}

\usepackage[utf8]{inputenc} 
\usepackage[T1]{fontenc}    
\usepackage{amsmath}
\usepackage{hyperref}       
\usepackage{url}            
\usepackage{booktabs}       
\usepackage{amsfonts}       
\usepackage{nicefrac}       
\usepackage{microtype}      
\usepackage[dvipsnames]{xcolor}         
\newcommand{\gain}[1]{{\scriptsize\color{ForestGreen}$\uparrow #1$}}
\newcommand{\loss}[1]{{\scriptsize\color{red}$\downarrow #1$}}

\usepackage{graphicx}       
\usepackage{xspace}
\usepackage{multirow}
\usepackage{tikz}
\usepackage{subcaption}
\usetikzlibrary{arrows.meta, decorations.pathmorphing, decorations.markings, positioning, fadings, shapes.geometric, calc}

\newcommand{\kt}{\ensuremath{k_{\mathrm{T}}}\xspace}
\newcommand{\fastjet}{\textsc{FastJet}\xspace}

\newcommand{\ttbar}{\ensuremath{\mathrm{t}\overline{\mathrm{t}}}\xspace}

\newcommand{\toptagging}{{\textsc{Top Quark Tagging}}\xspace}
\newcommand{\jetnet}{{\textsc{JetNet}}\xspace}
\newcommand{\jetclass}{{\textsc{JetClass}}\xspace}

\newcommand{\clstoken}{\texttt{[CLS]}}
\newcommand{\ie}{\textit{i.e.},\xspace}
\newcommand{\eg}{\textit{e.g.},\xspace}

\newcommand{\githubrepo}{\url{https://github.com/zichunhao/RINO}}

\title{RINO: Renormalization Group Distillation with No Labels}

\iclrfinalcopy
\author{%
  Zichun Hao, Raghav Kansal\thanks{Also affiliated with the Fermi National Accelerator Laboratory, Batavia, IL 60510, USA}\hspace{0.4em}\thanks{Now at Bexorg, Inc.}\hspace{0.45em}, Chang Sun, Maria Spiropulu
  \\
  Division of Physics, Mathematics and Astronomy \\
  California Institute of Technology \\
  Pasadena, CA 91125 \\
  \texttt{\{zhao,rkansal,chsun,smaria\}@caltech.edu} \\
  \\
  \AND
  Abhijith Gandrakota, Jennifer Ngadiuba \\
  Physics Division \\
  Fermi National Accelerator Laboratory \\
  Batavia, IL 60510 \\
  \texttt{\{abhijith,ngadiuba\}@fnal.gov} \\
  \And
  Haoyang Li, Zihan Zhao, Javier Duarte \\
  Department of Physics \\
  University of California San Diego \\
  La Jolla, CA 92093 \\
  \texttt{\{hal113,ziz078,jduarte\}@ucsd.edu}
}

\fancypagestyle{firstpage}{\rhead{\small FERMILAB-CONF-26-0756-CSAID-PPD}}
\begin{document}

\maketitle
\thispagestyle{firstpage}

\begin{abstract}
  Supervised machine learning in high-energy physics (HEP) requires training on labeled simulations of particle collisions, which are known to be unfaithful to the underlying physics.
  This leads to significant inaccuracies and biases in the transfer of trained models to real data, limiting the power of deep learning methods.
  Self-supervised learning (SSL) is a promising alternative for pretraining directly on real, unlabeled data.
  However, existing techniques largely operate with inductive biases inspired by domains such as images and natural language.
  We propose RINO (\textbf{R}enormalization Group D\textbf{I}stillation with \textbf{NO} Labels),
  which, for the first time, exploits the fundamentally multi-scale renormalization group structure of high-energy \textit{jet} formation as a physically grounded source of SSL views.
  We define teacher-student representations across energy scales using different levels of hierarchical clustering, aligned via DINO-style self-distillation.
  To mirror the background-dominated nature of real collider data, RINO pretrains exclusively on jets from quantum chromodynamics (QCD) processes and is evaluated on out-of-distribution (OOD) transfer to rare signal classes.
  In our controlled data-efficiency study, RINO matches fully supervised training on \jetnet{} after finetuning on only 1\% of the labeled data.
  With the full labeled dataset, RINO improves OOD classification accuracy on top-quark jets by 5.9\% and, without retraining, improves AUC on the unseen Higgs boson decay channel $H \to 4q$ by 4.1\% over the strongest SSL baseline.
  Scale-structure diagnostics reveal that DINO-style self-distillation permits our backbone to retain scale-dependent structure.
  Ablation studies show that this accounts for an 8.6\% OOD accuracy gap with respect to alternative SSL paradigms such as SimCLR.
  Code is available at \githubrepo{}.
\end{abstract}

\section{Introduction}
\label{sec:intro}
The discrepancy between simulated training data and real measurements is a fundamental bottleneck for machine learning (ML) in the physical sciences.
In high-energy physics (HEP), particles produced in real collisions cannot be labeled with absolute certainty, so supervised classifiers are trained using Monte Carlo (MC) simulations with well-defined collision contents (\ie labels).
However, due to the resulting domain shift, trained models can lose 20\% to 40\% of their performance when applied to real data~\citep{CMS-BTV-22-001,CMS-DP-2025-010}.
This gap stems from imperfect modeling of effects such as nonperturbative quantum chromodynamics (QCD) and detector response.
Self-supervised learning (SSL) offers a way to reduce the reliance on simulated data by pretraining directly on unlabeled real data and finetuning on labeled samples~\citep{SSL-robustness,SSL-single-image,SSL-survey}.
To do so, most SSL methods construct augmented \emph{views} of the same input and train the model to treat them as equivalent, with the choice of views implicitly determining the inductive bias of the learned representation.
A number of physics-motivated SSL methods have been developed for HEP~\citep{JetCLR,Zhao:2024kry,MPMv1,MPMv2,omnijet-alpha,J-JEPA,HEP-JEPA}, utilizing pretext tasks such as learning to align augmented views based on symmetries~\citep{JetCLR}, predicting masked particle constituents~\citep{MPMv1,MPMv2}, and joint-embedding predictions of held-out subjets or constituents~\citep{J-JEPA,HEP-JEPA}.
These designs encode valuable symmetries; however, they all operate on a single physical scale, ignoring an important structural property of the input.

A central object of study in HEP is the \emph{jet}: a collimated spray of particles produced by quark and gluon emission at high-energy colliders such as the CERN Large Hadron Collider (LHC)~\citep{Andersson:1987pr}.
From an ML perspective, a jet is a variable-length point cloud, while from a physics perspective, it is the observable end of QCD radiation, a stochastic cascade whose kinematic structure is \emph{multi-scale} by construction.
Identifying the source of a jet, called ``jet tagging'', is itself a well-studied ML task with a rich body of supervised methods~\citep{EFN,interaction-networks,Moreno_2020,ParticleNet,ParT,LorentzNet,PELICAN,L-GATr,Lorentz-equivariant-transformer-LHC,hepmllivingreview}.
Concretely, the same jet can be described at different resolutions: as a fine-grained cloud of $\mathcal{O}$(30--100) particles; a coarse-grained set of 2--4 subjets that capture only the high-level substructure; or anywhere in between.
These descriptions correspond to physically meaningful coarse-graining of the same underlying object, effectively the renormalization group (RG) flow from the high-energy collision scale down to the hadronization scale~\citep{PhysRevD.105.054012}.
RG structure has been widely studied across the physical sciences, such as cosmology~\citep{Pietroni:2011iz,Ivanov:2022eft,Calzetta:1991xe}, materials science~\citep{CG-protein,Wang:2019cgmd,ingolfsson:2014CG}, and fluid dynamics~\citep{Frisch:1995turbulence,Grinstein:2024turbulence,Berendsen_2007}, suggesting that methods designed around physical coarse-graining may generalize beyond HEP.


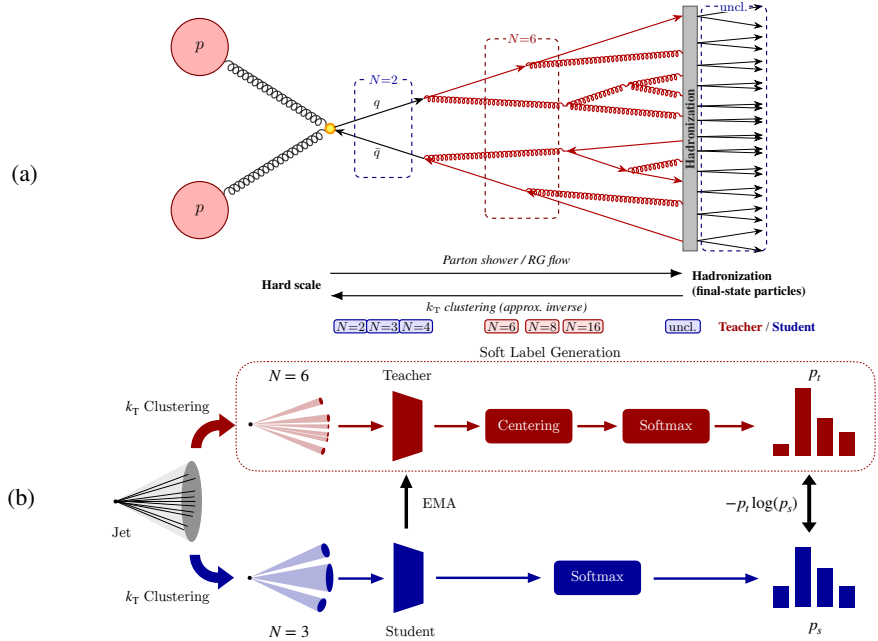
\begin{figure}[t]
  \centering
  \begin{subfigure}[c]{0.9\linewidth}
    \begin{minipage}[c]{0.03\linewidth}
      \subcaption{}\label{fig:parton-shower}
    \end{minipage}%
    \begin{minipage}[c]{0.96\linewidth}
      \centering
      \resizebox{0.72\linewidth}{!}{

\tikzset{
    gluon/.style={decorate, decoration={coil, aspect=0.7, segment length=3pt, amplitude=2pt}, draw=red!70!black, thick},
    gluonbig/.style={decorate, decoration={coil, aspect=0.9, segment length=4pt, amplitude=3pt}, draw=gray!40!black, thick},
    quark/.style={-{Stealth[length=2.5mm]}, thick, draw=black},
    quarkred/.style={-{Stealth[length=2.5mm]}, thick, draw=red!70!black},
    antiquark/.style={{Stealth[length=2.5mm]}-, thick, draw=black},
    antiquarkred/.style={{Stealth[length=2.5mm]}-, thick, draw=red!70!black},
    hadron/.style={-{Stealth[length=2mm]}, thin, draw=black},
    proton/.style={circle, draw=red!50!black, fill=red!30, minimum size=1.4cm, thick},
    qball/.style={circle, draw=black, fill=#1, minimum size=5mm, inner sep=0pt, font=\scriptsize\bfseries},
    splitbox/.style={draw=black, thick, rounded corners=2pt, minimum width=4.2cm, minimum height=1.8cm},
}

\begin{tikzpicture}[scale=1.0, every node/.style={font=\small}]


\node[proton] (p1) at (0, 2.0) {\large $p$};
\node[proton] (p2) at (0, -2.0) {\large $p$};

\coordinate (vtx) at (3.2, 0);
\draw[gluonbig] (p1) -- (vtx);
\draw[gluonbig] (p2) -- (vtx);

\fill[orange!80!yellow] (vtx) circle (4pt);
\fill[yellow!80] (vtx) circle (2.5pt);

\coordinate (qstart) at (5.5, 0.736);
\coordinate (qbarstart) at (5.5, -0.736);
\draw[quark, shorten <=3pt]     (vtx) -- (qstart)    node[midway, above, font=\small] {$q$};
\draw[antiquark, shorten <=3pt] (vtx) -- (qbarstart) node[midway, below, font=\small] {$\bar q$};

\fill[black!25] (11.85, -3.0) rectangle (12.20, 3.0);
\draw[thick, black!60] (11.85, -3.0) rectangle (12.20, 3.0);
\node[rotate=90, font=\footnotesize\bfseries, black!85] at (12.025, 0) {Hadronization};

\coordinate (a1q) at (8.0,  1.536);    
\coordinate (b1q) at (11.85, 2.768);   

\coordinate (a1g)   at (9.0,   0.55);   
\coordinate (b1g)   at (11.85, 1.85);   

\coordinate (a1g_a) at (10.5, 1.05);    
\coordinate (a1g_b) at (11.85, 0.30);   

\coordinate (a1g_a1) at (11.85, 1.30);  
\coordinate (a1g_a2) at (11.85, 0.80);  

\draw[quarkred] (qstart) -- (a1q);
\draw[quarkred] (a1q)    -- (b1q);
\draw[gluon]    (qstart) -- (a1g);
\draw[gluon]    (a1q)    -- (b1g);
\draw[gluon]    (a1g)    -- (a1g_a);
\draw[gluon]    (a1g)    -- (a1g_b);
\draw[gluon]    (a1g_a)  -- (a1g_a1);
\draw[gluon]    (a1g_a)  -- (a1g_a2);

\coordinate (la1q) at (8.0, -1.536);
\coordinate (lb1q) at (11.85, -2.768);

\coordinate (la1g)   at (9.0,   -0.55);
\coordinate (lb1g)   at (11.85, -1.85);

\coordinate (la1g_a)  at (10.5, -1.05);    
\coordinate (la1g_b)  at (11.85, -0.30);   

\coordinate (la1g_a_q) at (11.85, -1.30);  
\coordinate (la1g_a_g) at (11.85, -0.80);  

\draw[antiquarkred] (qbarstart) -- (la1q);
\draw[antiquarkred] (la1q)      -- (lb1q);
\draw[gluon]        (qbarstart) -- (la1g);
\draw[gluon]        (la1q)      -- (lb1g);
\draw[quarkred]     (la1g)      -- (la1g_a);     
\draw[antiquarkred] (la1g)      -- (la1g_b);     
\draw[quarkred]     (la1g_a)    -- (la1g_a_q);   
\draw[gluon]        (la1g_a)    -- (la1g_a_g);   

\foreach \y/\dy in { 2.75/0.25,  2.10/0.15,  1.55/0.10,
                      1.05/0.07,  0.55/0.05,  0.20/0.04 } {
    \draw[hadron] (12.20, \y) -- (13.80, \y + \dy);
    \draw[hadron] (12.20, \y) -- (13.80, \y - \dy);
}
\foreach \y/\dy in {-2.75/0.25, -2.10/0.15, -1.55/0.10,
                     -1.05/0.07, -0.55/0.05, -0.20/0.04 } {
    \draw[hadron] (12.20, \y) -- (13.80, \y + \dy);
    \draw[hadron] (12.20, \y) -- (13.80, \y - \dy);
}

\coordinate (sfL) at (3.2, -3.55);
\coordinate (sfR) at (11.85, -3.55);

\draw[thick, -{Latex[length=3mm]}] (sfL) -- (sfR)
    node[midway, above=1pt, font=\small\itshape] {Parton shower / RG flow};

\draw[thick, {Latex[length=3mm]}-] ($(sfL) + (0,-0.55)$) -- ($(sfR) + (0,-0.55)$)
    node[midway, below=1pt, font=\small\itshape] {$k_\mathrm{T}$ clustering (approx.\ inverse)};

\node[anchor=east, font=\small\bfseries] at ($(sfL) + (-0.1,-0.275)$) {Hard scale};
\node[anchor=west, font=\small\bfseries, align=left] at ($(sfR) + (0.1,-0.275)$) {Hadronization\\(final-state particles)};

\def\methrow{-4.85}

\foreach \x/\Nval in {3.7/2, 4.5/3, 5.3/4}{
  \node[draw=blue!60!black, fill=blue!12, rounded corners=2pt, thick,
        font=\footnotesize, inner sep=1.5pt, minimum width=0.65cm]
    at (\x, \methrow) {$N{=}\Nval$};
}

\foreach \x/\Nval in {7.4/6, 8.4/8, 9.4/16}{
  \node[draw=red!60!black, fill=red!12, rounded corners=2pt, thick,
        font=\footnotesize, inner sep=1.5pt, minimum width=0.65cm]
    at (\x, \methrow) {$N{=}\Nval$};
}

\node[draw=blue!60!black, fill=blue!12, rounded corners=2pt, thick,
      font=\footnotesize, inner sep=1.5pt, minimum width=0.85cm]
  at (11.85, \methrow) {$\mathrm{uncl.}$};


\draw[dashed, thick, blue!50!black, rounded corners=4pt]
  (3.8, -1.2) rectangle (5.2, 1.2);
\node[font=\footnotesize\bfseries, blue!60!black, fill=white, inner sep=1pt] at (4.5, 1.2) {$N{=}2$};

\draw[dashed, thick, red!50!black, rounded corners=4pt]
  (7.0, -2.2) rectangle (8.8, 2.2);
\node[font=\footnotesize\bfseries, red!60!black, fill=white, inner sep=1pt] at (7.9, 2.2) {$N{=}6$};

\draw[dashed, thick, blue!50!black, rounded corners=4pt]
  (12.3, -3.0) rectangle (13.9, 3.0);
\node[font=\footnotesize\bfseries, blue!60!black, fill=white, inner sep=1pt] at (13.1, 3.0) {$\mathrm{uncl.}$};

\node[anchor=west, font=\footnotesize, black!70] at (12.6, \methrow)
  {\textcolor{red!60!black}{\textbf{Teacher}} / \textcolor{blue!60!black}{\textbf{Student}}};

\end{tikzpicture}}
    \end{minipage}
  \end{subfigure}
  \\[2pt]
  \begin{subfigure}[c]{0.9\linewidth}
    \begin{minipage}[c]{0.03\linewidth}
      \subcaption{}\label{fig:rino-pipeline}
    \end{minipage}%
    \begin{minipage}[c]{0.96\linewidth}
      \centering
      \includegraphics[width=0.85\linewidth]{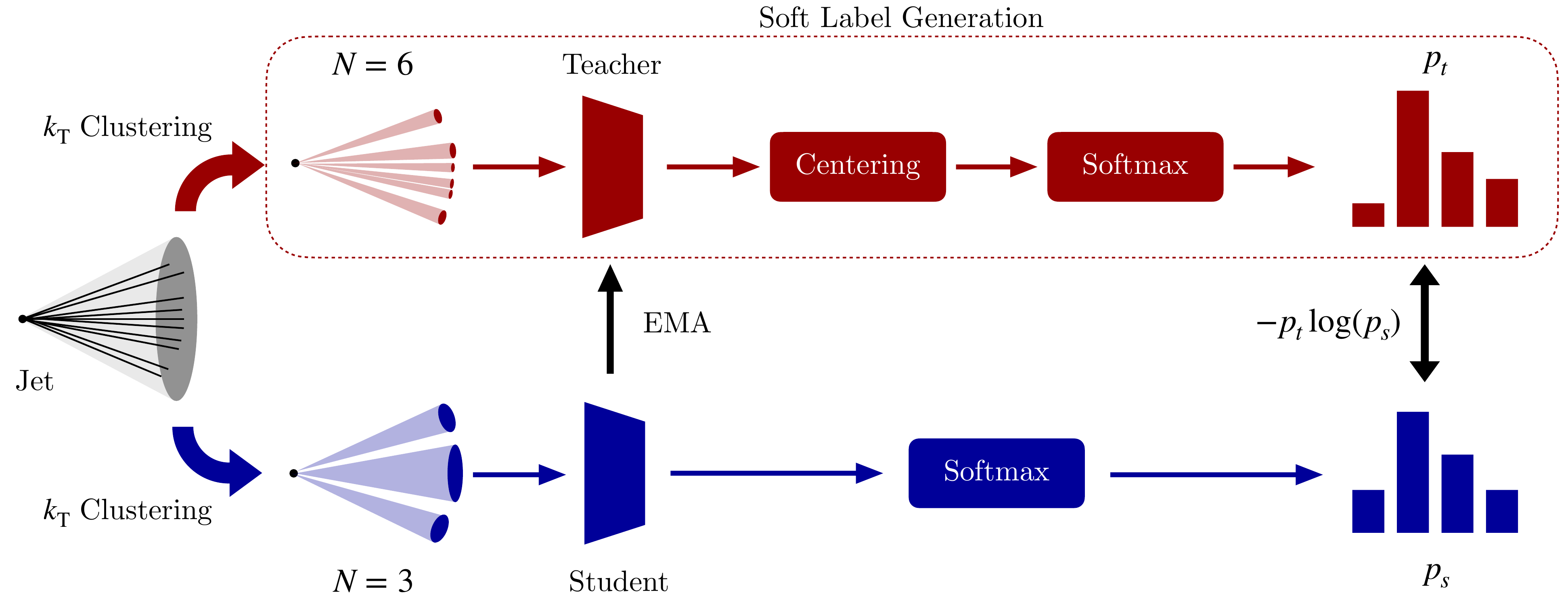}
    \end{minipage}
  \end{subfigure}
  \vspace{-4pt}
  \caption{\textbf{(a)} A QCD parton shower (here, a $q\bar q$ pair radiating softer gluons and quarks) is analogous to a renormalization group (RG) flow down to the hadronization scale. The \kt{} clustering approximately inverts this process, so stopping at different numbers of subjets $N$ generates the multi-scale views used for RINO self-distillation. \textbf{(b)} RINO training pipeline (DINO term of Eq.~\ref{eq:rino-loss}). The same jet is clustered at two \kt{} stopping points (here teacher $N{=}6$, student $N{=}3$). The teacher produces soft labels for the student to match via cross-entropy. Teacher parameters are updated from the student via an exponential moving average (EMA). This figure shows one representative student-teacher pair; in practice the loss aligns each student view to every teacher view at a different scale (Sec.~\ref{sec:method:views}).}
  \label{fig:overview}
  \vspace{-4pt}
\end{figure}

To exploit this fundamental multi-scale property of jets, we propose RINO (\textbf{R}enormalization Group D\textbf{I}stillation with \textbf{NO} Labels), an SSL framework for learning rich and robust jet representations.
RINO uses teacher-student self-distillation~\citep{DINO,DINOv2,DINOv3,self-distillations} over jet views generated at different energy scales by a hierarchical clustering algorithm called \textit{\kt{} clustering}~\citep{kt-clustering}.
Training across these scales encourages representations aligned across resolutions, while retaining scale-structured geometry in the backbone.

Our contributions are the following:
\begin{itemize}
  \item \textbf{A physics-informed multi-scale SSL framework.} To the best of our knowledge, RINO is the first SSL framework to align representations of a jet across multiple \kt{} clustering scales.
  \item \textbf{A realistic controlled comparison.} RINO is pretrained exclusively on QCD jets, the dominant background process at hadron colliders, in contrast to prior HEP SSL work that pretrains on simulated mixtures with signal classes overrepresented relative to real data. To probe transfer to deployment-like conditions, we finetune on the smaller \jetnet{} dataset and evaluate out-of-distribution (OOD) on the larger \jetclass{}, where \jetnet{} stands in for limited labeled simulation data and \jetclass{} for the larger target distribution, with a distribution shift between the two.
  This protocol finetunes without target-domain labels, matching the realistic downstream task.
  Under this protocol, RINO outperforms baselines on OOD top tagging by 5.9\% over the strongest baseline with lower seed-to-seed variance, and exhibits ${\sim}3\times$ smaller 1-Wasserstein distance ($W_1$) and 6--7$\times$ smaller squared maximum mean discrepancy (MMD$^2$) between the finetuning and evaluation domains.
  \item \textbf{Ablation studies identifying both DINO and scale-based distillation as core improvements.} We observe that both DINO and SimCLR~\citep{SimCLR} achieve comparable cross-scale invariance as a whole system (\ie backbone + projection head), but differ in where. DINO delegates it to the projection head while retaining scale-dependent structure in the backbone, whereas the backbone of our JetCLR-scale model is itself nearly scale-invariant. This difference is associated with an $8.6\%$ OOD accuracy gap. We further find that the multi-scale view construction contributes $+11.4\%$ to $+14.0\%$ over single-scale alternatives (Fig.~\ref{fig:factorization}).
\end{itemize}

\section{Related Work}
\label{sec:related-work}

\paragraph{Jet Tagging.}
Jet tagging has been the focus of numerous ML advances using deep sets, graph networks, transformers, and equivariant architectures~\citep{EFN,interaction-networks,ParticleNet,ParT,LorentzNet,PELICAN,L-GATr,Lorentz-equivariant-transformer-LHC,hepmllivingreview}.
These advances are largely supervised and architecture-driven, and therefore depend on labeled MC simulations whose imperfect modeling of detector and physics effects creates a persistent gap when the trained models are deployed on real collision data.
RINO instead focuses on the pretraining signal that produces transferable jet representations, learning directly from unlabeled data and using jet tagging only as its primary downstream benchmark.

\paragraph{Learning from Augmented Views.}
Several SSL methods learn representations by aligning differently augmented views of the same input~\citep{SimCLR,MOCO,SwAV,VICReg,BarlowTwins}.
DINO~\citep{DINO} advances this via teacher-student self-distillation~\citep{self-distillations}, and DINOv2~\citep{DINOv2} adds iBOT-style~\citep{iBOT} masked prediction on patch tokens.
In computer vision, views come from random cropping and photometric augmentation of dense pixel grids.
However, these augmentations do not transfer directly to jets, which are variable-length point clouds with no fixed spatial topology, so the choice of view generation shapes the inductive bias.

\paragraph{SSL in HEP.}
There has been growing interest in applying SSL techniques to HEP, spanning contrastive~\citep{JetCLR}, masked~\citep{MPMv1,MPMv2}, generative~\citep{omnijet-alpha}, and joint-embedding paradigms~\citep{J-JEPA,HEP-JEPA}.
JetCLR~\citep{JetCLR,Zhao:2024kry} uses SimCLR-style contrastive learning with physics-motivated augmentations such as rotations, translations, and soft and collinear augmentations.
MPMv1~\citep{MPMv1} tokenizes jet constituents via a VQ-VAE~\citep{VQVAE} and predicts discrete token IDs, in analogy with masked language modeling~\citep{BERT}, while MPMv2~\citep{MPMv2} reconstructs masked constituent features directly.
OmniJet-$\alpha$~\citep{omnijet-alpha} trains an autoregressive decoder on VQ-VAE tokens of jet constituents.
J-JEPA~\citep{J-JEPA} and HEP-JEPA~\citep{HEP-JEPA} adapt the JEPA framework~\citep{JEPA} to jets, predicting the embeddings of held-out subjets or constituents from the remaining ones.
In parallel, jBOT~\citep{jBOT} adapts iBOT to jets and evaluates in-distribution on \jetnet{}.
Other approaches include supervised pretraining on auxiliary physics tasks~\citep{Li:2024htp,Mikuni:2024qsr,Mikuni:2025tar,Mokhtar:2025zqs} and learning from simulation resampling~\citep{Resimulation}.
While these approaches introduce valuable domain-specific adaptations, they generate views at a single physical resolution.
RINO constructs views from the unclustered particle cloud down to a few subjets via \kt{} clustering and evaluates transfer from labeled simulation to a distinct target, mirroring on-data use.

\paragraph{Multi-Scale and Hierarchical SSL.}
Multi-scale view construction is a recurring theme in vision SSL.
DINO~\citep{DINO} combines large global crops with small local crops to encourage local-global consistency, and hierarchical vision transformers such as Swin~\citep{swin} build features at multiple resolutions.
These approaches define scale spatially (through pixel support) or architecturally, relying on artificial augmentations as a proxy for the underlying generative process.
By contrast, RINO defines scale through the QCD-motivated \kt{} clustering sequence: the number of remaining subjets indexes a physically ordered sequence of resolution scales.
The scale axis is therefore inherited from the physics that produces the data, and we analyze the resulting scale structure in Sec.~\ref{sec:disc:why}.


\paragraph{RG Flow and Jet Clustering.}
A parton shower is a stochastic realization of a QCD renormalization-group flow, in which partons from the hard scattering successively branch into lower-scale partons until the evolution reaches the hadronization scale~\citep{Nagy:2017ggp}.
Jet clustering algorithms approximately invert this process by sequentially combining nearby constituents.
At each step, the \kt{} algorithm~\citep{kt-clustering} merges the pair that minimizes
\begin{equation}
  d_{ij} = \min\!\left(p_{\mathrm{T}i}^2,\,p_{\mathrm{T}j}^2\right)\,\frac{\Delta R_{ij}^2}{R^2},
  \label{eq:kt-merge}
\end{equation}
or removes constituent $i$ if its beam distance $d_{iB} = p_{\mathrm{T}i}^2$ is smallest,
where $\Delta R_{ij}^2 = (\Delta\eta_{ij})^2 + (\Delta\phi_{ij})^2$ and $R$ is the jet radius parameter, so soft and collinear pairs are merged first to mimic the reverse of QCD splitting~\citep{Salam:2010nqg,Thaler:2010tr,CMS:2025eyd,jet-substructure-LHC,jet-substructure-LHC-future-colliders}.
Stopping at $N$ subjets yields a coarse-grained jet at a characteristic scale (App.~\ref{appendix:kt-scales};~\citealp{PhysRevD.105.054012}), and varying $N$ traces a physically ordered multi-scale axis for the same jet.
We use ``RG flow'' as a heuristic analogy: \kt{} clustering provides a sequence of coarse-grained jet descriptions ordered by an emission scale.
RINO converts this coarse-graining sequence into an SSL strategy by aligning representations \textit{across} clustering scales.

\section{Method: The RINO Framework}
\label{sec:method}
\subsection{Pretraining and Finetuning Data}
\label{sec:method:data}
We pretrain RINO exclusively on the QCD-jet subset of \jetclass~\citep{JetClass}, excluding the simulated signal classes (\eg top quark, $W$ boson, $Z$ boson, and Higgs boson jets).
While prior HEP SSL work pretrains on simulated mixtures of signal classes, real collider data is overwhelmingly composed of QCD jets.
Hence, restricting pretraining to QCD jets is more realistic for real-data SSL in HEP.
This also allows downstream evaluation to measure generalization to rare signal classes unseen during pretraining.
Meanwhile, \jetclass{} provides ground-truth labels for the held-out signal classes used in downstream evaluation, which would otherwise require signature-dependent triggers and physics cuts on real data that could themselves introduce biases.

For finetuning, we use the smaller \jetnet{}~\citep{JetNet-dataset} as a proxy for limited labeled simulation data and evaluate OOD on the held-out signal classes of \jetclass{}, the larger target distribution.
The two datasets differ in $p_\mathrm{T}$ range, constituent count, and detector simulation, inducing a distribution shift between finetuning and evaluation (App.~\ref{appendix:data-details}).

\subsection{Jet Clustering as Multi-Scale View Construction}
\label{sec:method:views}
RINO follows DINO's self-distillation framework~\citep{DINO,DINOv2,DINOv3}, but generates physics-informed views, each of which is a kinematic description of the same jet at a different scale.
We construct these views by stopping the exclusive \kt{} clustering sequence at varying numbers of subjets $N$, which integrates out soft and collinear radiation below a corresponding resolution scale (Sec.~\ref{sec:related-work}, Fig.~\ref{fig:parton-shower}).
Varying $N$ therefore produces a sequence of views of the same jet at physically ordered resolution scales.
We use the discrete ladder $N \in \{2, 3, 4, 6, 8, 16\}$ in addition to the fully unclustered particle cloud (denoted $\mathrm{uncl.}$), with $N \in \{2,3,4\}$ covering the typical hard prong multiplicities at the LHC.
Clustering histories are precomputed with \fastjet~\citep{FastJet} and cached.

During training, the teacher and student receive asymmetric view sets.
In the DINO term, the teacher sees only the intermediate-scale views ($N \in \{6, 8, 16\}$), while the student additionally sees the views at the two extremes of the ladder: coarse-grained ($N \in \{2, 3, 4\}$) and unclustered ($\mathrm{uncl.}$).
The DINO loss aligns each student view's representation with the teacher's intermediate-scale representations, encouraging multi-scale substructure encoding rather than single-scale specialization.
We ablate this choice in App.~\ref{appendix:ablation}.

\subsection{Setup: Training Objective and Architecture}
\label{sec:method:setup}
We adopt the composite pretraining objective of DINOv2~\citep{DINOv2} (Fig.~\ref{fig:rino-pipeline}), with the teacher parameters tracking the student via an exponential moving average (EMA)~\citep{EMA,MeanTeacher,MOCO}.
The RINO objective is
\begin{equation}
  \mathcal{L}
  = \lambda_\mathrm{DINO}\mathcal{L}_\mathrm{DINO}
  + \lambda_\mathrm{iBOT}\mathcal{L}_\mathrm{iBOT}
  + \lambda_\mathrm{KoLeo}\mathcal{L}_\mathrm{KoLeo}.
  \label{eq:rino-loss}
\end{equation}
The first term, $\mathcal{L}_\mathrm{DINO}$, aligns representations across the view sets (Sec.~\ref{sec:method:views}) along the parton shower hierarchy, with Sinkhorn--Knopp centering~\citep{SinkhornKnopp,SwAV} to prevent collapse.
The second term, $\mathcal{L}_\mathrm{iBOT}$, provides local supervision via masked-particle prediction: the student predicts the teacher's representation of each particle at masked positions. This serves a dual role: a single-scale masked-prediction baseline analogous to MPM-style methods~\citep{MPMv1,MPMv2} that we ablate against in Sec.~\ref{sec:exp:ablation}, and an auxiliary co-training objective that sharpens representation quality alongside DINO (App.~\ref{appendix:ablation}).
The third term,  $\mathcal{L}_\mathrm{KoLeo}$, is the Kozachenko--Leonenko (KoLeo) differential entropy estimator~\citep{KoLeo-Beirlant,KoLeo-Delattre,KoLeo-Sablayrolles}, which regularizes the embeddings for diversity on the hypersphere.
Appendices~\ref{appendix:pretrain-hparams} and~\ref{appendix:loss-defs} give the hyperparameters and the explicit loss terms.

Each jet is a point cloud of particle momentum vectors (an $N_\mathrm{p} \times 3$ tensor) preprocessed into $7$ kinematic features following the Particle Transformer (ParT)~\citep{ParT}, fed to a shared student-teacher transformer encoder~\citep{transformer}.
A polar-binned positional encoding (PE) indexed by each constituent's $(\Delta\eta, \Delta\phi)$ position breaks the degeneracy among masked iBOT tokens while preserving permutation equivariance.
Architecture details are provided in App.~\ref{appendix:arch-details}.

\section{Experiments}
\label{sec:exp}
\subsection{Setup and Controlled Comparison Protocol}
\label{sec:exp:setup}
All comparisons follow a controlled protocol, a simulated proxy for the intended workflow on experimental data (App.~\ref{appendix:deployment}), designed to isolate the contribution of the pretraining strategy.
Methods share the same jet transformer encoder backbone, except for the decoder-based OmniJet-$\alpha$ baseline, where we add causal masking to support its GPT-style task (App.~\ref{appendix:baselines}).
SSL methods are finetuned with the LP-FT~\citep{LP-FT} strategy, while the supervised baseline is trained end-to-end from random initialization. All methods use the same pretraining data (namely the \jetclass{} QCD class, a proxy for QCD-dominated real data), downstream data splits, finetuning hyperparameters (optimizer, learning-rate scheduler, weight decay; App.~\ref{appendix:finetune-protocol}), and number of finetuning seeds.
We evaluate top tagging (hadronic $t \to bqq$ vs.\ QCD jets) under the \jetnet{}$\to$\jetclass{} OOD protocol of Sec.~\ref{sec:method:data}, with binary classification accuracy on the \jetclass{} test set as the primary metric.
For each pretrained model, we report two downstream-evaluation levels, each with two metrics.
On the one hand, we probe the embedding quality by directly running a $k$-nearest-neighbors ($k$-NN; $k=20$) classifier and a linear probe on the frozen pretrained backbone after re-introducing \jetclass{} top jets and labels.\footnote{We would like to point out that such labels are unavailable in deployment on real, unlabeled data, so these two metrics serve only to demonstrate and diagnose embedding quality.}
On the other hand, a linear classification head and a two-layer MLP head are finetuned together with the backbone on \jetnet{} labels and evaluated on \jetclass{}'s top-tagging task to measure OOD generalization.\footnote{In this case, \jetclass{} top jets and all \jetclass{} labels (QCD and top) are not used in pretraining or finetuning.}
This hierarchy disentangles representation quality from OOD generalization.
Unless otherwise noted, numbers are reported as mean $\pm$ standard deviation over $10$ finetuning seeds; for pretrained backbones, the $k$-NN and linear-probe uncertainties are over probe fits on disjoint data subsets.

\subsection{OOD Top Tagging}
\label{sec:exp:main}
Table~\ref{tab:main-comparison} summarizes top-tagging accuracy on \jetclass{} for RINO, four recent SSL baselines (JetCLR, MPMv1, MPMv2, and OmniJet-$\alpha$), and supervised training from scratch on \jetnet{}.
To further isolate the contribution of multi-scale view construction, we include JetCLR-scale, a controlled variant that replaces JetCLR's augmentations with the same clustering-scale views used by RINO while keeping the contrastive objective (see App.~\ref{appendix:baselines} for details).

RINO is the only method whose MLP head significantly exceeds its linear head ($t=+3.0$, $p\approx 0.01$); no baseline shows the same improvement, and JetCLR and MPMv2 in particular show significant degradation ($t<-3$; App.~\ref{appendix:mlp-linear}).
This indicates that RINO's representations are robust enough that the additional capacity of the MLP head improves, rather than degrades, OOD accuracy.
With the MLP head, RINO demonstrates a $5.9\%$ relative improvement over the strongest baseline (MPMv1) and $7.8\%$ over the supervised model trained on \jetnet{}.
We also find that replacing JetCLR's augmentations with \kt{} views (JetCLR-scale) improves accuracy on the two OOD metrics, demonstrating the benefits of our scale-based augmentations under domain shift.
RINO also has the lowest seed-to-seed variance with the MLP head ($\pm 0.003$, versus at least $\pm 0.016$ for every other method), and its worst accuracy over five weight-decay and dropout settings ($0.843$) exceeds the best of supervised training, MPMv1, and MPMv2 (App.~\ref{app:finetune-robustness}), demonstrating robustness to common regularization hyperparameters (weight decay and dropout).
When the domain shift is smaller (\toptagging{}$\to$\jetclass{}), the leading methods saturate near $0.93$ in accuracy (App.~\ref{appendix:toptagging}), yet RINO's proxy scale factors stay closest to unity (App.~\ref{appendix:sf-proxy}).

\begin{table}[h]
  \centering
  \caption{Top-tagging accuracy on \jetclass{} (top vs.\ QCD). The linear and MLP heads are finetuned on \jetnet{} only (OOD); $k$-NN and the linear probe are fit on \jetclass{} labels and measure frozen-embedding quality.
    The bottom block groups all methods using \kt{}-clustering views (ours), differing only in the training loss.
    Each entry is $\mu \pm \sigma$; bold marks the best per column; \underline{underline} marks entries within $1\sigma$ of the best.}
  \label{tab:main-comparison}
  \small
  \setlength{\tabcolsep}{5pt}
  \vspace{-4pt}
  \begin{tabular}{lcccc}
    \toprule
    \textbf{Method}                       & $k$-NN                     & Linear probe                  & Linear head                & MLP head                   \\
    \midrule
    Supervised                            & $0.722 \pm 0.022$          & $0.714 \pm 0.030$             & $0.784 \pm 0.033$          & $0.786 \pm 0.023$          \\
    \midrule
    \multicolumn{5}{l}{\textit{SSL baselines}}                                                                                                                   \\
    JetCLR~\citep{JetCLR}                  & $0.780 \pm 0.002$          & $0.793 \pm 0.001$             & $0.791 \pm 0.048$          & $0.700 \pm 0.031$          \\
    OmniJet-$\alpha$~\citep{omnijet-alpha} & $0.859 \pm 0.001$          & $0.869 \pm 0.002$             & $0.749 \pm 0.080$          & $0.751 \pm 0.050$          \\
    MPMv1~\citep{MPMv1}                    & $0.882 \pm 0.002$          & $0.907 \pm 0.002$             & $0.809 \pm 0.008$          & $0.800 \pm 0.019$          \\
    MPMv2~\citep{MPMv2}                    & $0.896 \pm 0.001$          & $0.912 \pm 0.001$             & $0.817 \pm 0.027$          & $0.760 \pm 0.016$          \\
    \midrule
    \multicolumn{5}{l}{\textit{\kt{}-view SSL (ours)}}                                                                                                           \\
    JetCLR-scale                          & $0.884 \pm 0.001$          & $\underline{0.915 \pm 0.001}$ & $0.803 \pm 0.017$          & $0.780 \pm 0.023$          \\
    \textbf{RINO}                         & $\mathbf{0.908 \pm 0.001}$ & $\mathbf{0.916 \pm 0.001}$    & $\mathbf{0.838 \pm 0.009}$ & $\mathbf{0.847 \pm 0.003}$ \\
    \bottomrule
  \end{tabular}
  \vspace{-4pt}
\end{table}

\subsection{Data Efficiency}
\label{sec:exp:data-efficiency}
\begin{figure}[h]
  \centering
  \includegraphics[width=0.55\linewidth]{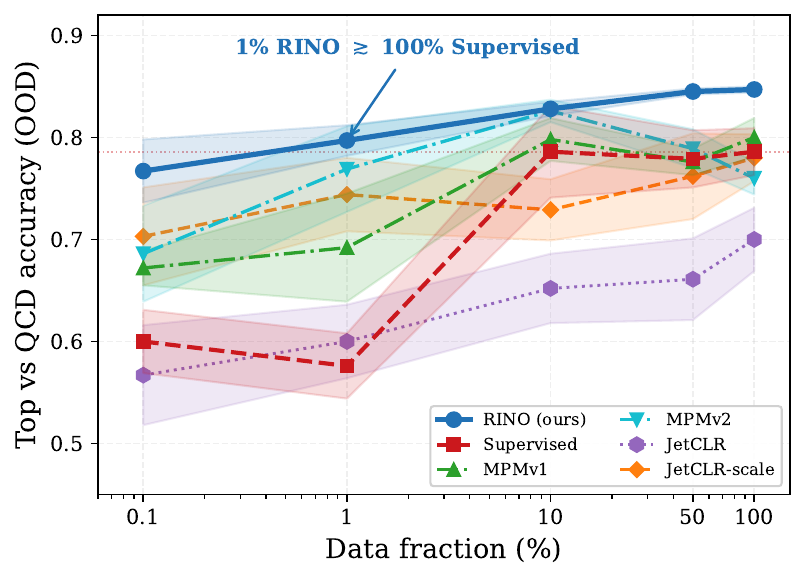}
  \vspace{-4pt}
  \caption{OOD top-tagging accuracy versus \jetnet{} data fraction.
    RINO with $1\%$ of labels matches supervised training on $100\%$ within uncertainty. Bands show $\pm1\sigma$ over $10$ finetuning seeds (MLP head); the dotted line marks supervised training on $100\%$ of labels.
  }
  \label{fig:data-efficiency}
  \vspace{-4pt}
\end{figure}

We sweep the \jetnet{} finetuning data fraction over $\{0.1, 1, 10, 50, 100\}\%$ and compare RINO with representative baselines, under the same finetuning protocol (App.~\ref{appendix:finetune-protocol}) to ensure the comparison reflects representational differences rather than tuning effort. Figure~\ref{fig:data-efficiency} reports OOD top-tagging accuracy on \jetclass{}.
RINO finetuned on $1\%$ of labels reaches $0.797 \pm 0.015$ accuracy, matching fully supervised training on the entire \jetnet{} dataset ($0.786 \pm 0.023$) within uncertainty.

RINO has the leading OOD accuracy at all data fractions and improves monotonically.
In contrast, some models are non-monotonic: MPMv2, for example, peaks in OOD performance at $10\%$ of the finetuning dataset.
This is consistent with the feature-distortion mechanism of \citet{LP-FT}: when pretrained features are strong and the OOD shift is large, finetuning distorts features along in-distribution (ID) directions.
More ID labels drive further distortion, which can explain this non-monotonic behavior. Moreover, JetCLR-scale outperforms JetCLR at all label fractions, so the benefit of cross-view alignment persists across data regimes. Full curves are in App.~\ref{app:data-efficiency}.

\subsection{Cross-Topology Generalization}
\label{sec:exp:multitask}
To test whether the learned representations capture general jet substructure rather than task- or dataset-specific features, we evaluate all methods on unseen \jetclass{} signal topologies after finetuning only on top tagging from \jetnet{} with a linear head (Table~\ref{tab:cross-topology}).
No model sees Higgs samples during any training.
We also report AUC as a metric, since the classification threshold is calibrated for the finetuning task and does not necessarily transfer to unseen signals and topologies.

RINO achieves the highest AUC on both $H \to 4q$ ($0.894 \pm 0.007$) and $H \to b\bar{b}$ ($0.788 \pm 0.016$).
The $H \to 4q$ result demonstrates genuine cross-topology generalization.
Although finetuned only on 3-prong top-quark jets (3 hard subjets from $t \to bqq$), RINO's linear classifier identifies 4-prong Higgs decays ($H \to 4q$) with low variance, demonstrating that \kt{}-view pretraining learns multi-prong substructure itself rather than only topology-specific features.

On $H \to b\bar{b}$, the leading methods cluster within ${\sim}2\%$ AUC of each other. This is expected: $H \to b\bar{b}$ discrimination relies on track-level features (impact parameters and secondary-vertex displacements) that distinguish b-quarks from light quarks. Our kinematic-only inputs cannot access these flavor signatures, so all methods are bounded by what kinematic substructure each backbone preserves.


\begin{table}[h]
  \centering
  \caption{Cross-topology transfer (\jetnet{}$\to$\jetclass{}, linear head).
    Models are finetuned on top tagging only and evaluated on unseen \jetclass{} signal topologies.
    Bold marks the best per column; \underline{underline} marks entries within one standard deviation of the best.}
  \label{tab:cross-topology}
  \small
  \setlength{\tabcolsep}{5pt}
  \vspace{-4pt}
  \begin{tabular}{lcccc}
    \toprule
    \textbf{Method}  & \multicolumn{2}{c}{$H \to 4q$} & \multicolumn{2}{c}{$H \to b\bar{b}$}                                                                 \\
    \cmidrule(lr){2-3} \cmidrule(lr){4-5}
                     & ACC                            & AUC                                  & ACC                           & AUC                           \\
    \midrule
    Supervised       & $0.707 \pm 0.026$              & $0.800 \pm 0.029$                    & $0.623 \pm 0.020$             & $0.726 \pm 0.030$             \\
    \midrule
    \multicolumn{5}{l}{\textit{SSL baselines}}                                                                                                               \\
    JetCLR~\citep{JetCLR}           & $0.727 \pm 0.067$              & $0.803 \pm 0.042$                    & $\mathbf{0.706 \pm 0.041}$    & $\underline{0.782 \pm 0.018}$ \\
    MPMv1~\citep{MPMv1}            & $0.741 \pm 0.027$              & $0.819 \pm 0.018$                    & $0.612 \pm 0.026$             & $0.705 \pm 0.017$             \\
    MPMv2~\citep{MPMv2}            & $0.757 \pm 0.033$              & $0.859 \pm 0.018$                    & $0.648 \pm 0.023$             & $\underline{0.778 \pm 0.010}$ \\
    OmniJet-$\alpha$~\citep{omnijet-alpha} & $0.692 \pm 0.053$              & $0.797 \pm 0.034$                    & $0.603 \pm 0.026$             & $0.709 \pm 0.032$             \\
    \midrule
    \multicolumn{5}{l}{\textit{\kt{}-view SSL (ours)}}                                                                                                       \\
    JetCLR-scale     & $0.630 \pm 0.020$              & $0.819 \pm 0.014$                    & $0.575 \pm 0.015$             & $0.771 \pm 0.010$             \\
    \textbf{RINO}    & $\mathbf{0.821 \pm 0.008}$     & $\mathbf{0.894 \pm 0.007}$           & $\underline{0.687 \pm 0.025}$ & $\mathbf{0.788 \pm 0.016}$    \\
    \bottomrule
  \end{tabular}
  \vspace{-4pt}
\end{table}

\subsection{Design Ablations}
\label{sec:exp:ablation}
\paragraph{Cross-scale DINO objective.}
Table~\ref{tab:ablation} progressively adds RINO components on top of an iBOT-only variant (single-scale masked prediction, no cross-scale DINO objective).
Adding DINO-based multi-scale distillation across \kt{} views yields $+14.0\%$, while a final teacher-schedule refinement adds another $+1.0\%$. Therefore, most of the improvement comes from the cross-scale DINO objective.
The full progressive decomposition, including view-assignment and PE choices, is in App.~\ref{appendix:ablation}.

\paragraph{Multi-scale view alignment.} 
JetCLR-scale and RINO use the same multi-scale views from \kt{} but differ in the training objective (Table~\ref{tab:main-comparison}).
JetCLR-scale outperforms the original JetCLR ($0.780$ vs.\ $0.700$, $+11.4\%$), confirming that \kt{} views account for most of the gain regardless of the SSL family.
The remaining gap to RINO ($+7.6\%$; $+8.6\%$ for the final model) reflects the benefit of self-distillation over contrastive alignment (discussed in detail in Sec.~\ref{sec:disc:why}).\footnote{We compare against the model before teacher tuning ($0.839$, Table~\ref{tab:ablation}) because it shares the pretraining recipe of the iBOT-only variant, whereas teacher tuning was applied only to the final model.}


\paragraph{Clustering algorithm.} 
To isolate the role of the energy-ordered \kt{} scale sequence, we train variants using Cambridge/Aachen (C/A) clustering~\citep{ca-clustering}, which groups by angular proximity, and anti-\kt{}~\citep{anti-kt-clustering}, which clusters the hardest constituents first.
Under the same training objective, C/A yields $0.796 \pm 0.012$ and anti-\kt{} yields $0.693 \pm 0.026$, compared to \kt{}'s $0.839 \pm 0.005$.
All three algorithms produce multi-scale views, but they differ in how they define resolution: \kt{} coarsens by energy scale (approximately reversing the parton shower), C/A by angular proximity, and anti-\kt{} by hardness.
The ranking \kt{} $>$ C/A $>$ anti-\kt{} is consistent with the finding that \kt{} subjets align more closely with the prongs from the hard process than C/A or anti-\kt{} subjets~\citep{CMS:2025eyd}, confirming that the SSL views help most when the coarse-graining axis matches the physical process that generated the data.

\begin{table}[h]
  \centering
  \caption{Ablation: contribution of DINO self-distillation.
    All variants share the same polar-binned PE, backbone, and finetuning protocol. Arrows give the relative gain over the previous row.}
  \label{tab:ablation}
  \small
  \vspace{-4pt}
  \begin{tabular}{ll}
    \toprule
    \textbf{Configuration}          & \textbf{OOD top ACC}            \\
    \midrule
    iBOT only                       & $0.736 \pm 0.028$               \\
    + Cross-scale self-distillation & $0.839 \pm 0.005$ \gain{14.0\%} \\
    + Teacher HP tuning             & $0.847 \pm 0.003$ \gain{1.0\%}  \\
    \bottomrule
  \end{tabular}
  \vspace{-4pt}
\end{table}

\section{Discussion}
\label{sec:discussion}

\subsection{Why Cross-Scale Training Helps}
\label{sec:disc:why}
\begin{figure}[h]
  \centering
  \resizebox{0.7\linewidth}{!}{\begin{tikzpicture}[
        node distance=2.0cm and 4.5cm,
        every node/.style={align=center},
        box/.style={draw, rounded corners=3pt, minimum width=2.5cm, minimum height=0.9cm, font=\large},
        arrow/.style={->, thick, >=stealth},
    ]

    \node[box] (jetclr) {JetCLR\\$0.700$};
    \node[box, right=of jetclr] (jetclr-scale) {JetCLR-scale\\$0.780$};
    \node[box, below=of jetclr] (ibot) {iBOT-only\\$0.736$};
    \node[box, below=of jetclr-scale] (rino) {\textbf{RINO}\\$\mathbf{0.839}$};

    \draw[arrow, color=blue!70!black] (jetclr) -- node[above, font=\normalsize] {\kt{} scale views} node[below, font=\normalsize, color=blue!70!black] {$+11.4\%$} (jetclr-scale);
    \draw[arrow, color=blue!70!black] (ibot) -- node[above, font=\normalsize] {\kt{} scale views} node[below, font=\normalsize, color=blue!70!black] {$+14.0\%$} (rino);

    \draw[arrow, color=red!70!black] (jetclr) -- node[left, font=\normalsize] {self-distillation} node[right, font=\normalsize, color=red!70!black] {$+5.1\%$} (ibot);
    \draw[arrow, color=red!70!black] (jetclr-scale) -- node[left, font=\normalsize] {self-distillation} node[right, font=\normalsize, color=red!70!black] {$+7.6\%$} (rino);

    \node[above=0.2cm of jetclr, font=\normalsize\itshape, anchor=south] {Single-scale};
    \node[above=0.2cm of jetclr-scale, font=\normalsize\itshape, anchor=south] {Multi-scale (\kt{})};
    \node[left=0.8cm of jetclr, font=\normalsize\itshape, anchor=east] (lbl-contrastive) {\large Contrastive};
    \node[left=0.8cm of ibot, font=\normalsize\itshape, anchor=east] (lbl-distill) {\large Self-distillation};

    \node[right=0.8cm of jetclr-scale, font=\normalsize, text opacity=0] {Self-distillation};
    \node[right=0.8cm of rino, font=\normalsize, text opacity=0] {Self-distillation};

\end{tikzpicture}}
  \vspace{-4pt}
  \caption{Factorization of RINO's OOD accuracy gain.
    The top horizontal arrow replaces JetCLR augmentations with \kt{} scale views; the bottom horizontal arrow adds DINO-based multi-scale distillation across \kt{} views. Vertical arrows compare contrastive and self-distillation objectives. RINO is shown before teacher tuning ($0.839$, Table~\ref{tab:ablation}); the final model reaches $0.847$ (Table~\ref{tab:main-comparison}).}
  \label{fig:factorization}
  \vspace{-4pt}
\end{figure}
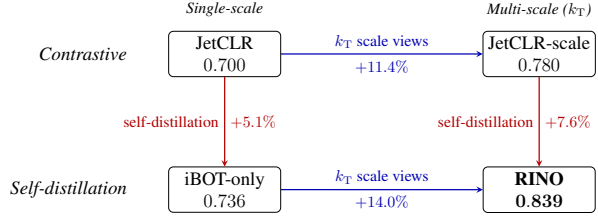

Two controlled comparisons isolate the contributions of \kt{} views and the SSL objective, as shown in Fig.~\ref{fig:factorization}.
Replacing JetCLR's augmentations with \kt{} views while keeping the contrastive loss (JetCLR-scale) improves accuracy from $0.700$ to $0.780$ ($+11.4\%$).
Adding cross-scale distillation over \kt{} views to the iBOT-only variant gives $+14.0\%$ ($0.736 \to 0.839$).
Switching from contrastive alignment to self-distillation adds a further $+5.1\%$ (JetCLR $\to$ iBOT-only) and $+7.6\%$ (JetCLR-scale $\to$ RINO before teacher tuning, $0.839$; Table~\ref{tab:ablation}).

Perhaps surprisingly, the mean cosine similarity across \kt{} clustering scales (App.~\ref{sec:exp:rg-invariance}, Fig.~\ref{fig:scale-similarity}) shows that RINO's backbone is \textit{not} scale-invariant, while the entire model (backbone and head combined) is one of the \textit{most invariant}.
Mean cosine similarity alone does not tell the whole story, though, since low similarity can also be inherited from input differences; the structure of the similarity matrix matters.
Unlike the scale-ignorant MPMv2 backbone (monotonic decay with distance along the \kt{} ladder) and the scale-invariant JetCLR-scale backbone (all cosines near $1$), RINO's backbone is scale-aware: it shows a block structure that follows the student-teacher role assignment.
This is permitted by projection-head shielding~\citep{Guillotine}: the loss only constrains $h \circ f$, so the backbone $f$ is free to develop any geometry the projection head $h$ can absorb.
RINO's invariance is therefore non-trivial: the head realizes invariance on top of a scale-aware backbone.
By contrast, the JetCLR-scale backbone is itself nearly scale-invariant (all cosines $\geq 0.89$). For OOD top tagging, the $8.6\%$ accuracy gap between RINO and JetCLR-scale (same views, same architecture) suggests that scale-aware backbones outperform scale-invariant ones.
Preserving scale information matters: for example, the top-tagging discriminative signal \emph{is} how jet substructure varies with scale. Top-quark jets resolve into three prongs at $N\!\sim\!3$ and develop characteristic soft-radiation patterns at relatively fine scales, whereas QCD jets do neither.

\subsection{Why RINO Generalizes OOD}
\label{sec:disc:ood-efficiency}
To measure the \jetnet{}$\to$\jetclass{} domain shift in representation space, we compute the 1-Wasserstein distance ($W_1$) and squared maximum mean discrepancy (MMD$^2$) between \jetnet{} and \jetclass{} finetuned embeddings, separately for QCD and top-quark jets (implementation details in App.~\ref{appendix:domain-shift}).
Both metrics quantify how far source-domain finetuned representations sit from the target distribution. Table~\ref{tab:domain-shift} shows that RINO has the smallest $W_1$ and MMD$^2$ of all methods ($W_1{=}0.043$ and MMD$^2{=}0.007$ for QCD, versus at least $0.132$ and $0.045$ for the baselines), alongside the highest OOD accuracy.
RINO's shift is already small before finetuning and does not grow during finetuning (App.~\ref{appendix:domain-shift}).

\begin{table}[h]
  \centering
  \caption{Domain shift in finetuned representations (\jetnet{} vs.\ \jetclass{}). Lower $W_1$ and MMD$^2$ indicate more domain-invariant representations.}
  \label{tab:domain-shift}
  \small
  \setlength{\tabcolsep}{4pt}
  \vspace{-4pt}
  \begin{tabular}{lccccc}
    \toprule
    \textbf{Method}  & $W_1$ (QCD) $\downarrow$ & $W_1$ (top) $\downarrow$ & MMD$^2$ (QCD) $\downarrow$ & MMD$^2$ (top) $\downarrow$ & OOD ACC $\uparrow$         \\
    \midrule
    Supervised       & $0.746$                  & $0.728$                  & $0.515$                & $0.564$                & $0.786 \pm 0.023$          \\
    JetCLR           & $0.132$                  & $0.127$                  & $0.196$                & $0.191$                & $0.700 \pm 0.031$          \\
    MPMv1            & $0.241$                  & $0.262$                  & $0.133$                & $0.162$                & $0.800 \pm 0.019$          \\
    MPMv2            & $0.375$                  & $0.348$                  & $0.338$                & $0.305$                & $0.760 \pm 0.016$          \\
    OmniJet-$\alpha$ & $0.815$                  & $0.752$                  & $0.571$                & $0.552$                & $0.751 \pm 0.050$          \\
    \midrule
    JetCLR-scale     & $0.133$                  & $0.145$                  & $0.045$                & $0.048$                & $0.780 \pm 0.023$          \\
    \textbf{RINO}    & $\mathbf{0.043}$         & $\mathbf{0.044}$         & $\mathbf{0.007}$       & $\mathbf{0.007}$       & $\mathbf{0.847 \pm 0.003}$ \\
    \bottomrule
  \end{tabular}
  \vspace{-4pt}
\end{table}

\subsection{Limitations}
First, our simulation-to-simulation proxy (\jetnet{}$\to$\jetclass{}) may not faithfully reflect the actual simulation-to-data shift. Appendix~\ref{appendix:deployment} outlines the intended workflow on experimental data, but direct real-data validation requires involved physics analyses (\eg control-region selection, tag-and-probe) and is beyond this work's scope.
Second, SSL methods pretrain on \jetclass{} QCD, so pretraining partially overlaps with the test domain; the genuine OOD challenge comes from unseen signal topologies and the mismatch in simulation settings between \jetnet{} and \jetclass{} QCD.
Third, the \kt{} inductive bias is task-dependent: RINO's advantage is largest for multi-prong substructure ($t \to bqq$, $H \to 4q$) and narrows for tasks driven by other signatures (\eg $H \to b\bar{b}$).

\section{Conclusion}
\label{sec:conclusion}
RINO constructs SSL views at multiple \kt{} clustering scales of the same jet and trains via DINO-style self-distillation. To simulate real-data conditions, we pretrain on \jetclass{} QCD only and finetune on the smaller labeled \jetnet{} before evaluating OOD on \jetclass{}. RINO outperforms baselines on OOD top tagging (Table~\ref{tab:main-comparison}), matches supervised training with $100\times$ fewer labels (Fig.~\ref{fig:data-efficiency}), and transfers without retraining to unseen $H \to 4q$ jets (Table~\ref{tab:cross-topology}).
A factorization study (Fig.~\ref{fig:factorization}) attributes this to both multi-scale view alignment and self-distillation, the latter preserving scale-aware backbone geometry (Sec.~\ref{sec:disc:why}).

\clearpage
\subsection*{AI use statement}
We used generative AI tools for the four purposes selected in the submission form, detailed below together with how the authors checked the results.

\paragraph{Aiding or polishing writing.}
We used generative AI tools to edit the paper for readability and to format references.
Every AI-suggested edit was shown to the authors as a diff and applied only after their review, modification, and approval.

\paragraph{Retrieval and discovery.}
We used generative AI tools to search for and identify references.
The authors required the tools to quote the supporting passage of each source verbatim, verified these quotes against the source, and checked that each citation supports the claim it is attached to. 
Every reference entry was checked against its \href{https://arxiv.org/}{arXiv}, \href{https://inspirehep.net/}{INSPIRE}, or DOI record.

\paragraph{Research ideation or execution.}
We used generative AI tools to:
\begin{itemize}\setlength{\itemsep}{0pt}
  \item execute and provide feedback on experiments;
  \item implement methods in a limited capacity, namely small fixes, bug fixes, and speed-only training optimizations to the authors' code in \githubrepo{}, adapting the open-source baseline implementations to our shared training and evaluation framework, and adapting JetCLR to the \kt{}-view JetCLR-scale variant;
  \item fix bugs in the data-preprocessing code;
  \item assist in interpreting results and in checking interpretations, including writing scripts to compute summary statistics (means and standard deviations);
  \item create and edit software code (\eg scripts, job configurations, and plotting code); and
  \item create and modify scientific figures.
\end{itemize}
The core RINO implementation and the preprocessing pipeline were written by the authors and were largely complete before generative AI tools were used in this project.
AI-assisted code, including plotting code, was reviewed and tested by the authors and validated through training and evaluation runs.
Derived numbers (\eg relative gains and test statistics) were recomputed with scripts rather than by hand and checked against the experiment outputs, and implementation details stated in the paper were checked against the code and configuration files.
Methodology and results were discussed in weekly group meetings, so that multiple authors reviewed them.
We have \emph{not} used generative AI tools to propose or refine hypotheses, develop theoretical models or conceptual frameworks, or assist with translation;
generating synthetic datasets, formulating or proving mathematical claims, and qualitative or thematic data analysis are not applicable to this work.

\paragraph{Drafting sections of the paper.}
We used generative AI tools to draft parts of the text, which the authors substantially rewrote.
Every drafted change was reviewed and approved by the authors before inclusion, and multiple authors then independently reviewed the full manuscript or its core parts.

\paragraph{Review and responsibility.}
The authors have reviewed all AI-assisted work as detailed above.
We take responsibility for the final content of this work, including text, claims, or artifacts produced with the aid of generative AI.

\subsection*{Ethics statement}
RINO aims to reduce reliance on MC simulation by enabling pretraining on real, unlabeled collider data, with no direct societal risks specific to this work.

\subsection*{Reproducibility statement}
Code is available at \githubrepo. Appendix~\ref{appendix:pretrain-hparams} lists the pretraining hyperparameters, App.~\ref{appendix:loss-defs} the loss terms, App.~\ref{appendix:arch-details} the architecture and input features, App.~\ref{appendix:data-details} the datasets and preprocessing, App.~\ref{appendix:finetune-protocol} the finetuning protocol, App.~\ref{appendix:baselines} the baseline implementations, and App.~\ref{appendix:compute} the compute resources. All datasets are public~\citep{JetClass,JetNet-dataset,TopTaggingDataset}.

\ificlrfinal
\subsubsection*{Acknowledgments}
  We thank Yue Ma and Aobo Li for thoughtful feedback on the experimental design and presentation.
  We are grateful to Joseph Lykken for valuable discussions and insightful comments that sharpened the narrative of this work.
  Z.H., R.K., C.S., and M.S. are supported by the California Institute of Technology High Energy Physics under Contract DE-SC0011925 with the United States Department of Energy, Office of High Energy Physics.
  Z.H. would like to thank the Fermilab LPC Guests and Visitors program.
  A.G., J.N., and J.D. are supported by the DOE Office of Science, Award No. DE-SC0023524, FermiForward Discovery Group, LLC under Contract No. 89243024CSC000002 with the U.S. Department of Energy, Office of Science, Office of High Energy Physics, LDRD L2024-066-1, Fermilab, DOE Office of Science, Office of High Energy Physics ``Designing efficient edge AI with physics phenomena'' Project (DE-FOA-0002705), DOE Office of Science, Office of Advanced Scientific Computing Research under the ``Real-time Data Reduction Codesign at the Extreme Edge for Science'' Project (DE-FOA-0002501).
  J.D. is also supported by the Research Corporation for Science Advancement (RCSA) under grant \#CS-CSA-2023-109, Alfred P. Sloan Foundation under grant \#FG-2023-20452, DOE, Office of Science, Office of High Energy Physics Early Career Research program under Award No. DE-SC0021187, DOE, Office of Science, Office of High Energy Physics under Award No. DE-SC0009919, and the U.S. National Science Foundation (NSF) Harnessing the Data Revolution (HDR) Institute for Accelerating AI Algorithms for Data Driven Discovery (A3D3) under Cooperative Agreement PHY-2117997.
  R.K. and J.N. are also supported by the AI2050 program at Schmidt Futures (Grant G-23-64934).
  This work was performed using the National Research Platform Nautilus HyperCluster supported by NSF awards CNS-1730158, ACI-1540112, ACI-1541349, OAC-1826967, OAC-2112167, CNS-2100237, CNS-2120019, the University of California Office of the President, and the University of California San Diego's California Institute for Telecommunications and Information Technology/Qualcomm Institute.
\fi

\clearpage

\bibliography{bibliography}
\bibliographystyle{iclr2027_conference}

\clearpage

\appendix

\section{Appendix}
The code is provided in \githubrepo.

\subsection{Pretraining Hyperparameters}
\label{appendix:pretrain-hparams}
Table~\ref{tab:pretrain-hparams} summarizes the hyperparameters used for RINO pretraining on the QCD-jet subset of \jetclass{}.
The teacher EMA momentum and the learning-rate scheduler follow cosine schedules over the full $100$ training epochs, with warmup as indicated.
The pretraining recipe was held fixed across all RINO ablation variants reported in Sec.~\ref{sec:exp:ablation}, with only the view-construction parameters and the input positional encoding varied between variants.
The headline RINO model in Table~\ref{tab:main-comparison} ($0.847 \pm 0.003$) uses a tuned teacher schedule, a fixed teacher temperature and a lower final EMA momentum (Table~\ref{tab:pretrain-hparams}), chosen by the OOD top-vs-QCD accuracy on \jetclass{}; this step contributes the final $+1.0\%$ row of Table~\ref{tab:ablation}, and RINO before tuning ($0.839 \pm 0.005$) already exceeds every baseline (App.~\ref{appendix:baselines}). Values listed in Table~\ref{tab:pretrain-hparams} below are those of this final configuration.

\begin{table}[h]
  \centering
  \caption{RINO pretraining hyperparameters (production configuration).}
  \label{tab:pretrain-hparams}
  \small
  \setlength{\tabcolsep}{5pt}
  \begin{tabular}{lll}
    \toprule
    \textbf{Group} & \textbf{Hyperparameter}      & \textbf{Value}                                                   \\
    \midrule
    \multirow{4}{*}{Optimization}
                   & Optimizer                    & AdamW ($\beta_1=0.9$, $\beta_2=0.999$)                           \\
                   & Peak learning rate           & $3\times10^{-3}$                                                 \\
                   & Weight decay                 & $0.05$                                                           \\
                   & Gradient clipping (max norm) & $3.0$                                                            \\
    \midrule
    \multirow{4}{*}{Schedule}
                   & Total epochs                 & $100$                                                            \\
                   & LR warmup                    & epochs $0$--$19$, $10^{-6} \to 1.0$ (cosine)                     \\
                   & LR main schedule             & epochs $20$--$99$, $1.0 \to 0.05$ (cosine)                       \\
                   & Layer-wise LR decay          & $0.9$ per layer (backbone)                                       \\
    \midrule
    \multirow{2}{*}{Batching}
                   & Total batch size             & $3000$ jets \\
                   & Per-GPU batch size           & $1500$ jets ($2$ GPUs, data parallel)                            \\
    \midrule
    \multirow{5}{*}{DINO loss}
                   & Student temperature $\tau_s$ & $0.10$ (fixed)                                                   \\
                   & Teacher temperature $\tau_t$ & $0.04$ (fixed)                                                   \\
                   & \quad ablation variants      & $0.04 \to 0.07$, epochs $0$--$19$ (cosine)                       \\
                   & Teacher centering            & Sinkhorn--Knopp                                                  \\
                   & Loss weight                  & $\lambda_\mathrm{DINO} = 1.0$                                    \\
    \midrule
    \multirow{3}{*}{iBOT loss}
                   & Mask ratio                   & $0.25$ (unclustered view)                                        \\
                   & Mask embedding               & learnable $d_\mathrm{model}$-dim vector                          \\
                   & Loss weight                  & $\lambda_\mathrm{iBOT} = 1.0$                                    \\
    \midrule
    \multirow{2}{*}{KoLeo regularizer}
                   & Applied to                   & student \clstoken{} embeddings (teacher-scale views)             \\
                   & Loss weight                  & $\lambda_\mathrm{KoLeo} = 0.1$                                   \\
    \midrule
    \multirow{3}{*}{Teacher EMA}
                   & Update rule                  & $\bar\theta \leftarrow m\,\bar\theta + (1-m)\,\theta$, per batch \\
                   & Momentum schedule            & $m: 0.995 \to 0.997$ over $100$ epochs (cosine)                  \\
                   & \quad ablation variants      & $m: 0.995 \to 0.999$ over $100$ epochs (cosine)                  \\
    \bottomrule
  \end{tabular}
\end{table}

\subsection{Pretraining Loss Terms}
\label{appendix:loss-defs}
This section writes out the three terms of Eq.~\ref{eq:rino-loss}; hyperparameter values are listed in Table~\ref{tab:pretrain-hparams}.
For a view $x$ of a jet, $P_s(x)$ and $P_t(x)$ denote the student and teacher output distributions over the $K$ prototypes of a projection head (App.~\ref{appendix:arch-details}), obtained with temperatures $\tau_s$ and $\tau_t$.
The teacher distribution is normalized with Sinkhorn--Knopp~\citep{SinkhornKnopp,SwAV} and receives no gradient.
We write the cross-entropy as
\begin{equation}
  H(a, b) = -\sum_{k=1}^{K} a_k \log b_k,
  \label{eq:cross-entropy}
\end{equation}
and each loss term below is averaged over the jets in a mini-batch.

\paragraph{DINO term.}
Let $x^{(v)}$ be the jet at view $v$, $\mathcal{T} = \{6, 8, 16\}$ the teacher views, and $\mathcal{S}$ the student views (the teacher views plus three student-only views sampled per batch from $\{2, 3, 4, \mathrm{uncl.}\}$).
Using the DINO head on the \clstoken{} output,
\begin{equation}
  \mathcal{L}_\mathrm{DINO}
  = \frac{1}{|\mathcal{P}|}\sum_{(t,s) \in \mathcal{P}} H\bigl(P_t(x^{(t)}),\, P_s(x^{(s)})\bigr),
  \qquad \mathcal{P} = \{(t, s) \in \mathcal{T} \times \mathcal{S} : s \neq t\},
  \label{eq:loss-dino}
\end{equation}
so each student view is aligned with every teacher view~\citep{DINO} at a different scale.

\paragraph{iBOT term.}
The iBOT term acts on the unclustered view $x$ only (App.~\ref{appendix:ablation}).
A random set $\mathcal{M}$ of $25\%$ of the particles is masked in the student's input, giving $\tilde x$, and the teacher sees the unmasked jet.
Using the iBOT head on the output token of each masked particle $i$~\citep{iBOT},
\begin{equation}
  \mathcal{L}_\mathrm{iBOT}
  = \frac{1}{|\mathcal{M}|}\sum_{i \in \mathcal{M}} H\bigl(P_t(x)_i,\, P_s(\tilde x)_i\bigr).
  \label{eq:loss-ibot}
\end{equation}

\paragraph{KoLeo term.}
Following DINOv2~\citep{DINOv2,KoLeo-Sablayrolles}, for the $\ell_2$-normalized student \clstoken{} representations $\hat z_1, \dots, \hat z_B$ of the $B$ jets in a mini-batch at the same teacher-scale view,
\begin{equation}
  \mathcal{L}_\mathrm{KoLeo}
  = -\frac{1}{B}\sum_{i=1}^{B} \log\Bigl(\min_{j \neq i}\,\lVert \hat z_i - \hat z_j \rVert_2\Bigr),
  \label{eq:loss-koleo}
\end{equation}
averaged over teacher-scale views. It spreads different jets within a view and never compares two views of the same jet.

\subsection{Architecture Details}
\label{appendix:arch-details}

\paragraph{Backbone.}
The shared transformer encoder backbone has $8$ layers with hidden dimension $d_\mathrm{model} = 256$, $16$ attention heads, and a feed-forward expansion ratio of $4$ (FFN hidden dimension $1024$), totaling $8.4$M trainable parameters (plus $2.6$M parameters in the polar-binned positional encoding lookup table).
Activations are SwiGLU~\citep{SwiGLU} in the FFN sub-layers, normalization is RMSNorm~\citep{RMSNorm} applied pre-residual, and residual paths use LayerScale~\citep{LayerScale} with learnable per-channel scaling initialized to $0.01$.
A learnable \clstoken{} token is prepended to each input sequence and serves as the global jet representation.
Following \citet{registers}, we additionally append four learnable \emph{register} tokens that are not used as outputs but stabilize the attention maps.
Variable-length particle sequences are handled with attention masking; the maximum sequence length is $90$ particles per jet (the $90$ highest-$p_\mathrm{T}$ particles are kept).

\paragraph{Input features (per particle).}
Each constituent is represented by the seven features listed in Table~\ref{tab:particle-features}.
All features are pre-standardized through a fixed affine map $x \mapsto (x - \mu) / \sigma$.
The per-feature centering $\mu$ and scale $\sigma$ are estimated once from the \emph{median} and \emph{interquartile range} (IQR) of the corresponding distribution on the \jetclass{} QCD pretraining set, rather than from the empirical mean and standard deviation; this robust estimator is insensitive to the long tails typical of jet kinematic spectra.
The resulting $(\mu, \sigma)$ values are held fixed during both training and inference, and used unchanged on all downstream datasets, so that the same backbone can be applied without renormalizing the input distribution.
Quantities quoted in $(\Delta\eta, \Delta\phi)$ space elsewhere in this appendix (notably the radial parameters of the positional encoding) are in these standardized units, so for example $\Delta\eta = 1$ corresponds to a raw rapidity offset of $0.14$.

\begin{table}[h]
  \centering
  \caption{Per-particle input features and the fixed standardization $(\mu, \sigma)$ applied by the input pipeline. The model receives $(x - \mu)/\sigma$ for each feature.}
  \label{tab:particle-features}
  \small
  \begin{tabular}{llr@{\quad}r}
    \toprule
    \textbf{Feature}                               & \textbf{Definition}                                                         & $\mu$  & $\sigma$ \\
    \midrule
    $\log p_\mathrm{T}$                            & $\log p_{\mathrm{T},i}$, constituent transverse momentum                    & $1.7$  & $1.8$    \\
    $\log E$                                       & $\log E_i$, constituent energy                                              & $2.0$  & $1.8$    \\
    $\log(p_\mathrm{T}/p_\mathrm{T}^\mathrm{jet})$ & $\log(p_{\mathrm{T},i}/p_\mathrm{T}^\mathrm{jet})$, relative $p_\mathrm{T}$ & $-4.7$ & $1.8$    \\
    $\log(E/E^\mathrm{jet})$                       & $\log(E_i/E^\mathrm{jet})$, relative energy                                 & $-4.7$ & $1.8$    \\
    $\Delta\eta$                                   & $\eta_i - \eta_\mathrm{jet}$, rapidity offset                               & $0.0$  & $0.14$   \\
    $\Delta\phi$                                   & $\phi_i - \phi_\mathrm{jet}$ (wrapped to $[-\pi,\pi)$), azimuthal offset    & $0.0$  & $0.14$   \\
    $\Delta R$                                     & $\sqrt{(\Delta\eta_i)^2 + (\Delta\phi_i)^2}$, angular distance to jet axis  & $0.14$ & $0.25$   \\
    \bottomrule
  \end{tabular}
\end{table}

\paragraph{Positional encoding.}
The polar-binned positional encoding maps each constituent's $(\Delta\eta, \Delta\phi)$ position to a learned embedding via a $100\!\times\!100$ angular grid (radial $\times$ azimuthal bins) with logarithmic radial spacing centered on the jet axis.
Concretely, the radial coordinate $r = \sqrt{(\Delta\eta)^2 + (\Delta\phi)^2}$ is mapped through $\log(1 + r / r_\mathrm{scale})$ before quantization into $100$ bins on the interval $[0,\, \log(1 + r_\mathrm{max}/r_\mathrm{scale})]$, while the azimuthal coordinate $\alpha = \operatorname{arctan2}(\Delta\phi, \Delta\eta)$ is binned uniformly on $[-\pi, \pi)$.
The two radial parameters are not free: $r_\mathrm{scale}$ is set to the median of the per-constituent radial distribution on the pretraining set (so half of all constituents fall in the high-resolution region near the knee of the log compression), and $r_\mathrm{max}$ is set to the 99th percentile of the same distribution (so extreme tail constituents are clamped rather than allowed to consume a large share of bins).
The two radial parameters are quoted in the standardized $(\Delta\eta, \Delta\phi)$ units of Table~\ref{tab:particle-features}; on the \jetclass{} QCD pretraining set this gives $r_\mathrm{scale} \approx 0.96$ and $r_\mathrm{max} \approx 5.60$, corresponding to ${\sim}0.13$ and ${\sim}0.78$ in raw $(\Delta\eta, \Delta\phi)$ space, within the $R = 0.8$ jet cone.
The combination of polar parameterization and log radial binning therefore concentrates resolution where the constituent density is highest, by construction.

Bin collisions matter because the role of the positional encoding in an encoder-only iBOT setup is precisely to break what would otherwise be a complete degeneracy among masked constituents.
With a generative decoder (as in MAE-style methods), positional queries fed to the decoder disambiguate which masked token to reconstruct.
With our encoder-only architecture, the only signal a masked token carries is its mask embedding plus its positional embedding; without PE, all masked tokens are identical and iBOT cannot distinguish them.
A bin collision restores this degeneracy locally: two masked constituents that fall in the same bin become indistinguishable.
Empirically, adding polar-binned PE helps with training stability and improves OOD accuracy from $0.828$ to $0.839$ on cluster views (Table~\ref{tab:ablation-studies}), consistent with reduced bin collisions aiding iBOT training.

We measure, for each candidate scheme, the per-particle collision rate (fraction of real constituents sharing a bin with at least one other constituent of the same jet) and the per-jet collision rate (fraction of jets containing at least one such collision).
The polar grid is markedly more bin-efficient than a Cartesian grid of comparable size because the constituent $(\Delta\eta, \Delta\phi)$ distribution is concentrated near the jet axis (the median jet has $32$ constituents, see App.~\ref{appendix:data-details}).
A representative comparison is shown in Table~\ref{tab:pe-collision}.

\begin{table}[h]
  \centering
  \caption{Bin-collision rates (\%) for positional-encoding schemes across grid sizes, measured on $5\!\times\!10^5$ \jetclass{} QCD jets. Lower collision rates reduce ambiguity among masked iBOT tokens. The production choice is Polar $100\!\times\!100$ (bold).}
  \label{tab:pe-collision}
  \begin{tabular}{lrrrrr}
    \toprule
    \textbf{Scheme}    & $20\!\times\!20$ & $50\!\times\!50$ & $80\!\times\!80$ & $\mathbf{100\!\times\!100}$ & $120\!\times\!120$ \\
    \textit{(patches)} & \textit{400}     & \textit{2{,}500} & \textit{6{,}400} & \textbf{\textit{10{,}000}}  & \textit{14{,}400}  \\
    \midrule
    \multicolumn{6}{l}{\textit{Per-particle collision rate (\%)}}                                                                  \\
    Cartesian          & 64.76            & 35.98            & 22.19            & 16.82                       & 13.13              \\
    Log-Cartesian      & 42.70            & 13.61            & 6.27             & 4.19                        & 3.00               \\
    \textbf{Polar}     & 32.96            & 9.87             & 4.52             & \textbf{3.03}               & 2.16               \\
    \midrule
    \multicolumn{6}{l}{\textit{Per-jet collision rate (\%)}}                                                                       \\
    Cartesian          & 99.93            & 99.44            & 97.49            & 94.55                       & 90.13              \\
    Log-Cartesian      & 99.13            & 83.55            & 59.55            & 46.79                       & 37.26              \\
    \textbf{Polar}     & 90.01            & 59.14            & 39.59            & \textbf{31.19}              & 24.94              \\
    \midrule
    \multicolumn{6}{l}{\textit{Mask--mask collision rate (\% of masked particles, $25\%$ mask)}}                                   \\
    Cartesian          & 38.37            & 14.71            & 7.42             & 5.17                        & 3.80               \\
    Log-Cartesian      & 16.99            & 3.82             & 1.59             & 1.02                        & 0.72               \\
    \textbf{Polar}     & 12.38            & 2.78             & 1.16             & \textbf{0.76}               & 0.53               \\
    \bottomrule
  \end{tabular}
\end{table}

Three observations emerge.
First, all three schemes improve monotonically with grid size, and the same ranking (Polar best, Log-Cartesian intermediate, Cartesian worst) holds at every grid size and on all three metrics.
Second, most of the improvement over uniform Cartesian comes from \emph{density-matched} binning (Log-Cartesian alone reduces the $100\!\times\!100$ per-particle rate from $16.82\%$ to $4.19\%$, a $4\times$ reduction); the polar parameterization adds a further $\sim\!1.4\times$ refinement ($4.19\% \to 3.03\%$) on top, by aligning the bin boundaries with the rotational symmetry of the jet.
Third, the mask--mask rate (bottom panel) is the operationally relevant figure for iBOT, and it is substantially smaller than the per-particle rate at every grid size: at the production Polar $100\!\times\!100$, only $0.76\%$ of masked particles share a bin with another masked particle (vs.\ $3.03\%$ per-particle overall), so the degeneracy iBOT actually faces during training is roughly $4\times$ lower than the full per-particle rate would suggest.

Polar $100\!\times\!100$ is a deliberate middle ground.
Going further to Polar $120\!\times\!120$ would reduce the per-particle collision rate only from $3.03\%$ to $2.16\%$ and the per-jet rate from $31.19\%$ to $24.94\%$, while requiring $44\%$ more patches and a correspondingly larger learnable PE table ($\sim\!1.13$M extra parameters at $d_\mathrm{model} = 256$).
Conversely, dropping to Polar $80\!\times\!80$ would save $36\%$ of the patches but increase the per-particle rate from $3.03\%$ to $4.52\%$.
The $100\!\times\!100$ choice therefore lies at the knee of the diminishing-returns curve, capturing nearly all of the achievable reduction without paying for the long tail of marginal improvements.

\paragraph{Projection heads.}
Following DINOv2~\citep{DINOv2}, the DINO and iBOT projection heads are independent three-layer MLPs with hidden width $1024$ and bottleneck dimension $256$, GELU activations, an $\ell_2$-normalization layer, and a weight-normalized linear output of dimension $2048$.
The DINO head operates on the \clstoken{} representation, whereas the iBOT head operates on per-particle outputs at masked positions.

\subsection{Dataset Details}
\label{appendix:data-details}


\paragraph{\jetclass.}
\jetclass~\citep{JetClass} is a large-scale jet classification benchmark with $10$ classes spanning QCD, Higgs, electroweak, and top topologies. The hard process is generated with \textsc{MadGraph5\_aMC@NLO}, parton showering and hadronization use \textsc{Pythia}, and detector effects are modeled with \textsc{Delphes} in the CMS configuration; jets are clustered from \textsc{Delphes} E-Flow objects with the anti-\kt{} algorithm at $R = 0.8$ and selected with $p_\mathrm{T} \in [500, 1000]$\,GeV and $|\eta| < 2$. See \citet{JetClass,ParT} for full details.
We use the QCD subset ($10$M training jets) for pretraining and the signal classes for downstream evaluation.

\begin{figure}[h]
  \centering
  \includegraphics[width=0.65\linewidth]{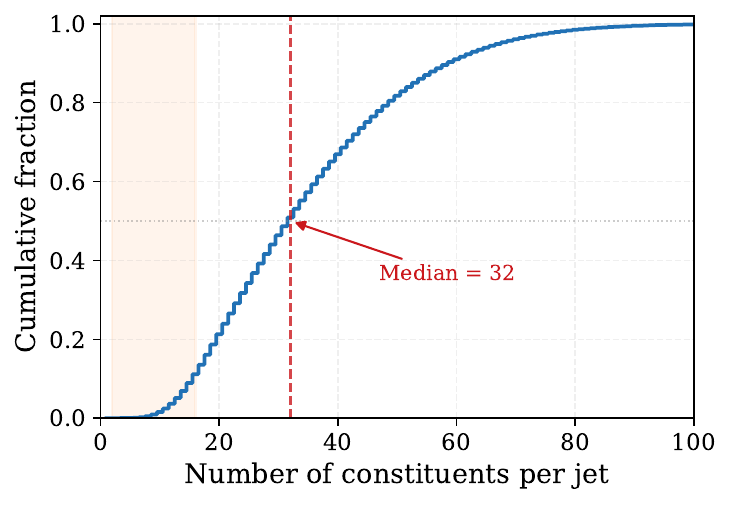}
  \caption{Cumulative distribution of constituent multiplicity for QCD jets in \jetclass{} ($12.5$M jets across train, validation, and test splits). The shaded region marks the \kt{} view range $N \in \{2, \ldots, 16\}$; only $11\%$ of jets have $\leq 16$ constituents, confirming that the coarse views are genuine coarse-grainings.}
  \label{fig:nparticles-cdf}
\end{figure}

\paragraph{Pretraining set (QCD only).}
We pretrain exclusively on the QCD subset ($10$M training jets), excluding all signal classes.
The median QCD jet contains $32$ constituents (mean $35.2 \pm 17.2$; Fig.~\ref{fig:nparticles-cdf}): only $11\%$ of jets have $\leq 16$ constituents and $0.13\%$ have $\leq 6$, confirming that the coarse views ($N \in \{2,3,4\}$) are genuine coarse-grainings for nearly all jets.
Subjet clustering is computed once during preprocessing with \fastjet~\citep{FastJet} using the exclusive \kt{} algorithm with jet radius $R = 0.8$, and the four-momenta of the resulting subjets are stored on disk for the stopping points $N \in \{2, 3, 4, 6, 8, 16\}$.

\paragraph{Downstream evaluation sets.}
We finetune on \jetnet~\citep{JetNet-dataset,JetNet-library} ($\sim\!360$K jets covering gluon, light-quark, and top topologies, with jet $p_\mathrm{T} \approx 1$\,TeV, restricted to $0.8$--$1.6$\,TeV, and at most $30$ constituents per jet, zero-padded). \jetnet{} parton-level events are generated with \textsc{MadGraph5\_aMC@NLO}, then showered and hadronized with \textsc{Pythia} (Monash 2013 tune), with a simplified detector-level description instead of \textsc{Delphes}. It therefore differs from \mbox{\jetclass{}} in $p_\mathrm{T}$ range and detector-level simulation.
We additionally use the \toptagging{} dataset~\citep{Kasieczka:2019dbj,TopTaggingDataset} ($\sim\!1.2$M training jets: top vs.\ QCD, with $p_\mathrm{T} \in [550, 650]$\,GeV and up to $200$ constituents), which uses \textsc{Pythia} for shower/hadronization and \textsc{Delphes} for detector effects with a different (ATLAS-like) configuration than \jetclass.
For OOD evaluation we test on the \jetclass{} test set, restricted to the relevant signal class versus QCD (\eg top vs.\ QCD for the main comparison).
All downstream evaluation samples are disjoint from the pretraining sample; the model never sees their labels during pretraining and never sees their non-QCD signal classes at all.
The \jetnet{}$\to$\jetclass{} transfer is OOD in three respects: different $p_\mathrm{T}$ range, different constituent count distribution (at most $30$ vs.\ up to $183$), and different detector-level simulation (simplified vs.\ \textsc{Delphes}).

\subsection{Physical Scales of the \kt{} Views}
\label{appendix:kt-scales}
The hierarchy is not learned but imposed by the number of subjets per view, $N$.
Each clustering step does indeed have a meaningful energy scale: the $N{+}1 \to N$ merge scale $q_N = \sqrt{d_\mathrm{merge}(N)}$~\citep{kt-clustering}, where $d_\mathrm{merge}(N)$ is the \kt{} distance ($d_{ij}$ of Eq.~\ref{eq:kt-merge} or $d_{iB}$) at the clustering step from $N{+}1$ to $N$ subjets.
For \jetclass{} QCD jets with more than $16$ constituents, we report the median and IQR of $q_N$ and of $p_\mathrm{T}^\mathrm{soft}/p_\mathrm{T}^\mathrm{jet}$, the transverse momentum of the softest of the $N$ subjets divided by that of the parent jet (Table~\ref{tab:kt-scales}).

\begin{table}[h]
  \centering
  \caption{Physical scales of the exclusive-\kt{} views: merge scale $q_N$ and softest-subjet momentum fraction, median [IQR] over \jetclass{} QCD jets with more than $16$ constituents.}
  \label{tab:kt-scales}
  \small
  \begin{tabular}{rcc}
    \toprule
    $N$  & $q_N$ [GeV]            & $p_\mathrm{T}^\mathrm{soft}/p_\mathrm{T}^\mathrm{jet}$ \\
    \midrule
    $2$  & $9.91\ [6.07, 17.09]$  & $0.23\ [0.12, 0.36]$       \\
    $3$  & $5.70\ [3.69, 9.01]$   & $0.088\ [0.042, 0.156]$    \\
    $4$  & $3.89\ [2.57, 5.99]$   & $0.044\ [0.021, 0.084]$    \\
    $6$  & $2.25\ [1.48, 3.39]$   & $0.018\ [0.009, 0.034]$    \\
    $8$  & $1.49\ [0.95, 2.26]$   & $0.010\ [0.006, 0.019]$    \\
    $16$ & $0.46\ [0.24, 0.76]$   & $0.0033\ [0.0021, 0.0054]$ \\
    \bottomrule
  \end{tabular}
\end{table}

The median scale $q_N$ falls monotonically from $9.91$ to $0.46$\,GeV, while the resolved softest-subjet fraction falls from $0.23$ to $0.0033$.
Thus, the fixed $N$ does indeed correspond to a meaningful distribution of physical scales.
Two results show that RINO benefits from this ordering.
First, the scale-structure analysis (App.~\ref{sec:exp:rg-invariance}) finds that the backbone retains resolution information while the projection head aligns across scales.
Second, with all other choices fixed, the intended teacher $\{6, 8, 16\}$/student-only $\{2, 3, 4, \mathrm{uncl.}\}$ assignment gives $0.839 \pm 0.005$, versus $0.823 \pm 0.006$ when the roles are swapped and $0.730 \pm 0.009$ when both roles mix scales (Table~\ref{tab:ablation-studies}).

\subsection{Finetuning Protocol}
\label{appendix:finetune-protocol}

Table~\ref{tab:finetune-hparams} gives the locked finetuning recipe used in all main results in Sec.~\ref{sec:exp}, identical across pretrained backbones. The supervised baseline shares the same optimizer, scheduler, and head architecture but is trained end-to-end from random initialization without the frozen-backbone warmup.
For SSL methods we adopt the LP-FT (linear-probe-then-finetune) strategy~\citep{LP-FT,ren2023preparetaskheadfinetuning}: the classification head is trained for the first 5 epochs with the backbone frozen, after which the backbone is unfrozen and trained jointly with the head at the same learning rate.
This was found to give substantially smaller seed-to-seed variance than full finetuning from epoch $0$, while matching its peak accuracy.
Accuracy uses a fixed decision threshold of $0.5$. The \jetclass{} test sets are class-balanced, while \jetnet{} has a QCD:top ratio of $2{:}1$, which the positive-class weight of $2.0$ compensates (Table~\ref{tab:finetune-hparams}).

\begin{table}[h]
  \centering
  \caption{Finetuning hyperparameters (locked across all backbones and datasets).}
  \label{tab:finetune-hparams}
  \small
  \begin{tabular}{ll}
    \toprule
    \textbf{Hyperparameter}                        & \textbf{Value}                                        \\
    \midrule
    Optimizer                                      & AdamW ($\beta_1=0.9$, $\beta_2=0.999$)                \\
    Head learning rate                             & $1\times10^{-4}$                                      \\
    Backbone learning-rate factor (after unfreeze) & $1.0$ (\ie same as head)                            \\
    Weight decay                                   & $0.05$                                                \\
    LP-FT: backbone freeze epochs                  & $5$                                                   \\
    LR warmup epochs                               & $10$                                                  \\
    LR scheduler                                   & cosine with warmup                                    \\
    Dropout (in head)                              & $0.1$                                                 \\
    Label smoothing                                & $0.0$                                                 \\
    Early-stopping patience                        & $5$ epochs (on validation loss)                       \\
    BCE positive-class weight                      & $2.0$ for \jetnet{}; $1.0$ for \toptagging{}          \\
    Head architecture                              & 2-layer MLP, hidden $[256, 128]$, ReLU, dropout $0.1$ \\
    Pooling                                        & mean over particle tokens (for backbones with PE)     \\
    Seeds per configuration                        & $10$ for \jetnet and $6$ for \toptagging              \\
    \bottomrule
  \end{tabular}
\end{table}

\subsection{Robustness to Finetuning Regularization}
\label{app:finetune-robustness}
All main results use one finetuning recipe for every method (Table~\ref{tab:finetune-hparams}).\footnote{The supervised baseline uses the same hyperparameters but is trained end-to-end, without the LP-FT frozen-backbone warmup (App.~\ref{appendix:finetune-protocol}).}
To test whether RINO's lead depends on this shared recipe, we vary the weight decay and dropout over five (weight decay, dropout) settings, $(0.01, 0.1)$, $(0.05, 0.1)$, $(0.10, 0.1)$, $(0.05, 0.3)$, and $(0.05, 0.5)$, for supervised training, MPMv1, MPMv2, and RINO with the default backbone and $100\%$ of the \jetnet{} labels, keeping all other settings of Table~\ref{tab:finetune-hparams}.
Table~\ref{tab:finetune-reg} reports the default setting (the Table~\ref{tab:main-comparison} value) together with the lowest (Worst) and highest (Best) mean OOD accuracy across the five settings.
Best is selected on \jetclass{} accuracy, which requires target labels that are unavailable in applications to real data, so it is a retrospective best case rather than a deployable choice.

For supervised training and MPMv1, no setting improves on the default, so the shared recipe does not disadvantage them.
RINO's worst setting ($0.843$) exceeds the best setting of every other method, and its default leads the strongest target-selected baseline (MPMv2, $0.810$) by $0.037$.
Its accuracy spans $0.009$ across the settings, comparable to MPMv1 ($0.010$) and smaller than supervised training ($0.015$) and MPMv2 ($0.050$).

\begin{table}[h]
  \centering
  \caption{Sensitivity of OOD top-tagging accuracy ($\mu \pm \sigma$ over $10$ finetuning seeds; \mbox{\jetnet{}}$\to$\jetclass{}, MLP head, $100\%$ of \jetnet{} labels) to the finetuning weight decay and dropout. Default is the Table~\ref{tab:main-comparison} value (weight decay $0.05$, dropout $0.1$); Worst and Best are the lowest and highest mean over the five settings of App.~\ref{app:finetune-robustness}. Best is selected on \jetclass{} accuracy and is a retrospective upper bound. Bold marks the best per column.}
  \label{tab:finetune-reg}
  \small
  \begin{tabular}{lccc}
    \toprule
    \textbf{Method}     & \textbf{Default}           & \textbf{Worst}             & \textbf{Best (target-selected)} \\
    \midrule
    Supervised          & $0.786 \pm 0.023$          & $0.771 \pm 0.033$          & $0.786 \pm 0.023$          \\
    MPMv1~\citep{MPMv1} & $0.800 \pm 0.019$          & $0.790 \pm 0.018$          & $0.800 \pm 0.019$          \\
    MPMv2~\citep{MPMv2} & $0.760 \pm 0.016$          & $0.760 \pm 0.016$          & $0.810 \pm 0.017$          \\
    \textbf{RINO}       & $\mathbf{0.847 \pm 0.003}$ & $\mathbf{0.843 \pm 0.005}$ & $\mathbf{0.852 \pm 0.003}$ \\
    \bottomrule
  \end{tabular}
\end{table}

\subsection{Extended Ablation Results}
\label{appendix:ablation}

Table~\ref{tab:ablation-progressive} traces the progressive construction of the RINO recipe.
Table~\ref{tab:ablation-studies} reports independent comparisons along the positional-encoding, view-assignment, and clustering-algorithm axes.
In the view sets, $\mathrm{uncl.}$ denotes the unclustered jet (all particles).
OOD ACC (accuracy) entries are mean $\pm$ standard deviation over $10$ seeds with an MLP head; the $k$-NN and linear-probe (LP) columns are the frozen-backbone diagnostics of Sec.~\ref{sec:exp:setup}.
The final model (row~5) uses a refined teacher schedule (fixed $\tau_t=0.04$, final EMA momentum $0.997$) compared to rows 1--4, which use a cosine warmup to $\tau_t=0.07$ and final momentum $0.999$.

\begin{table}[h]
  \centering
  \caption{Progressive construction of the RINO recipe. Each row builds on the previous row.}
  \label{tab:ablation-progressive}
  \small
  \begin{tabular}{llccl}
    \toprule
    \textbf{Step} & \textbf{Configuration}                            & $k$-NN  & LP      & OOD ACC                         \\
    \midrule
    1.            & DINO only, mixed views, no PE                     & $0.859$ & $0.923$ & $0.677 \pm 0.016$               \\
    2.            & + iBOT                                            & $0.897$ & $0.918$ & $0.776 \pm 0.018$ \gain{14.6\%} \\
    3.            & + \kt{} multi-scale views, $p_\mathrm{T}$-rank PE & $0.905$ & $0.916$ & $0.828 \pm 0.004$ \gain{6.7\%}  \\
    4.            & + Polar-binned PE                                 & $0.919$ & $0.919$ & $0.839 \pm 0.005$ \gain{1.3\%}  \\
    5.            & + Teacher HP tuning (production)                  & $0.908$ & $0.916$ & $0.847 \pm 0.003$ \gain{1.0\%}  \\
    \bottomrule
  \end{tabular}
\end{table}

\begin{table}[h]
  \centering
  \caption{Independent comparisons of design choices. Each block varies one axis while holding others fixed. Arrows give the relative change (\%) with respect to the first row of each block.}
  \label{tab:ablation-studies}
  \small
  \begin{tabular}{llccl}
    \toprule
    \textbf{Axis} & \textbf{Configuration}                                   & $k$-NN  & LP      & OOD ACC                      \\
    \midrule
    \multicolumn{5}{l}{\textit{PE choice $\mid$ mixed views, DINO+iBOT}}                                                        \\
                  & No PE                                                    & $0.897$ & $0.918$ & $0.776 \pm 0.018$            \\
                  & $p_\mathrm{T}$-rank                                      & $0.902$ & $0.919$ & $0.777 \pm 0.028$ \gain{0.1} \\
                  & Polar-binned                                             & $0.909$ & $0.922$ & $0.730 \pm 0.009$ \loss{5.9} \\
    \midrule
    \multicolumn{5}{l}{\textit{PE choice $\mid$ \kt{} multi-scale views, DINO+iBOT}}                                            \\
                  & $p_\mathrm{T}$-rank                                      & $0.905$ & $0.916$ & $0.828 \pm 0.004$            \\
                  & \textbf{Polar-binned}                                    & $0.919$ & $0.919$ & $0.839 \pm 0.005$ \gain{1.3} \\
    \midrule
    \multicolumn{5}{l}{\textit{View assignment $\mid$ polar-binned PE, DINO+iBOT}} \\
    \multicolumn{5}{l}{\footnotesize (sets shown are teacher views and student-only views; the student also sees the teacher's views)} \\
                  & \textbf{Teacher $\{6,8,16\}$, student $\{2,3,4,\mathrm{uncl.}\}$} & $0.919$ & $0.919$ & $0.839 \pm 0.005$            \\
                  & Swapped: teacher $\{2,3,4\}$, student $\{6,8,16,\mathrm{uncl.}\}$ & $0.903$ & $0.922$ & $0.823 \pm 0.006$ \loss{1.9} \\
                  & Mixed-scale: teacher $\{2,4,8,16\}$, student $\{3,6,\mathrm{uncl.}\}$ & $0.909$ & $0.922$ & $0.730 \pm 0.009$ \loss{13.0} \\
    \midrule
    \multicolumn{5}{l}{\textit{Clustering algorithm $\mid$ polar-binned PE, DINO+iBOT}}                                         \\
                  & \textbf{\kt{}}                                           & $0.919$ & $0.919$ & $0.839 \pm 0.005$            \\
                  & C/A                                                      & $0.878$ & $0.909$ & $0.796 \pm 0.012$ \loss{5.1} \\
                  & Anti-\kt{}                                               & $0.799$ & $0.906$ & $0.693 \pm 0.026$ \loss{17.4} \\
    \bottomrule
  \end{tabular}
\end{table}

\paragraph{Mixed-scale views.}
The ``mixed-scale'' view assignment reshuffles clustering scales between teacher and student so that both roles span the full range: the teacher sees $\{2, 4, 8, 16\}$ and the student additionally sees $\{3, 6, \mathrm{uncl.}\}$.
This preserves the use of multi-scale \kt{} views but breaks the hierarchy in which the teacher is restricted to resolved intermediate scales and the student-only views sit at both extremes (few subjets and unclustered).
The $13.0\%$ drop from hierarchical to mixed-scale views under the same DINO+iBOT loss (Table~\ref{tab:ablation-studies}) shows that the specific hierarchical assignment is the important design choice.

\paragraph{iBOT loss scale.}
In both RINO and the iBOT-only ablation, the iBOT loss is applied only on the full unclustered particle view. We do not train iBOT to reconstruct masked constituents in the $N$-subjet clustered views, so that the multi-scale part of RINO is isolated in the DINO objective.
Applying masked prediction across clustered views is a natural future extension.

\paragraph{DINO and iBOT are complementary.}
Neither $\mathcal{L}_\mathrm{DINO}$ nor $\mathcal{L}_\mathrm{iBOT}$ produces a strong representation alone: DINO-only (mixed views, no PE) reaches $0.677$ and iBOT-only reaches $0.736$ (Table~\ref{tab:ablation-progressive}, row 1; Table~\ref{tab:ablation}). Combined, they reach $0.776$ (mixed views) and $0.839$ (\kt{} views), exceeding what either component achieves on its own. $\mathcal{L}_\mathrm{iBOT}$ thus acts as an auxiliary co-training signal that sharpens representation quality when paired with $\mathcal{L}_\mathrm{DINO}$.

\paragraph{PE interaction with views.}
Polar-binned PE hurts on mixed views ($0.730$ vs.\ $0.776$ without PE) but helps on cluster views ($0.839$ vs.\ $0.828$ with $p_\mathrm{T}$-rank PE).
This interaction is specific to the view assignment: polar-binned PE encodes spatial position within each subjet view, and its benefit appears only when the teacher-student hierarchy is respected.
Under the mixed assignment, polar-binned PE degrades accuracy by $5.9\%$, while under the hierarchical assignment it improves accuracy by $1.3\%$.

\subsection{Scale Structure}
\label{sec:exp:rg-invariance}
\begin{figure}[!tbp]
  \centering
  \includegraphics[width=0.85\linewidth]{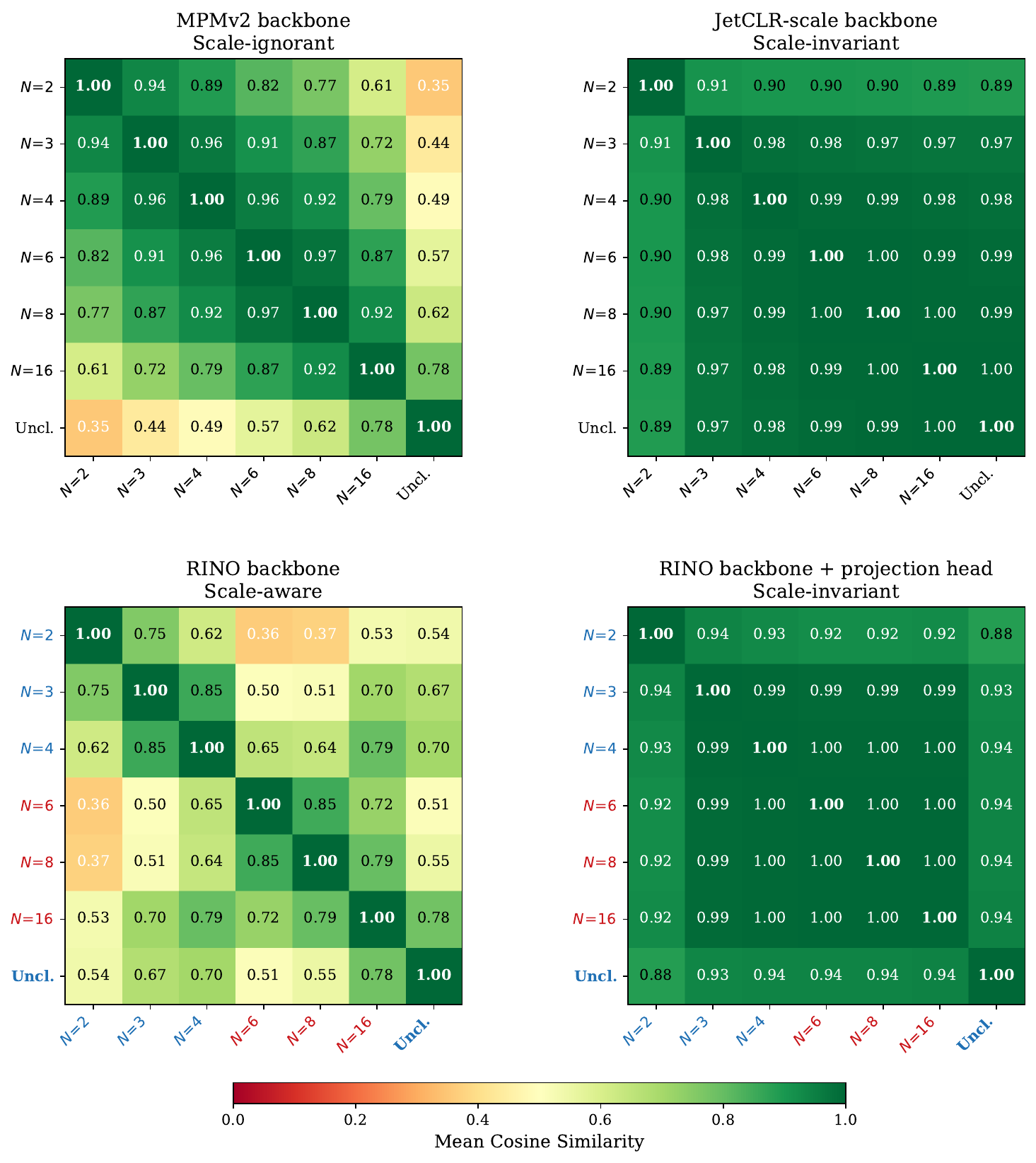}
  \caption{Per-jet mean cosine similarity across \kt{} clustering scales.
  Left to right, top to bottom: MPMv2 shows monotonic decay (input-similarity pattern); JetCLR-scale converges to near-invariance; RINO backbone retains scale-dependent geometry organized along the {\color[HTML]{2171b5}student}/{\color[HTML]{cb181d}teacher} role assignment; RINO backbone + projection head realizes cross-scale invariance in the $2048$-dimensional head output, including the unclustered view.}
  \label{fig:scale-similarity}
\end{figure}

A natural question is how models' representations encode the parton-shower hierarchy.
Figure~\ref{fig:scale-similarity} compares the per-jet mean cosine similarity matrix between clustering scales for three models.

A scale-ignorant baseline such as MPMv2 shows the pattern produced by input similarity alone, where, for a given clustering scale, cosine similarity decreases monotonically as the other scale moves away along the \kt{} ladder in either direction, ranging from $0.92$--$0.97$ between adjacent clustered scales to $0.35$ at the most distant pair ($N\!=\!2$ vs.\ the unclustered jet).
This monotonic decay reflects the trivial correlation between subjet count and input feature overlap, without any learned cross-scale structure.
In contrast, JetCLR-scale produces near-invariant representations across scales, with all cross-scale cosines in $[0.89, 1.00]$.

RINO shows a third, scale-aware pattern that matches neither baseline.
The $N\!=\!2$ row is particularly informative: $N\!=\!2$ is more similar to the unclustered view ($0.54$) and $N\!=\!16$ ($0.53$) than to the intermediate teacher-target scales $N\!=\!6$ ($0.36$) and $N\!=\!8$ ($0.37$), despite $N\!=\!6$ and $N\!=\!8$ being closer to $N\!=\!2$ in subjet count.
This non-monotonic structure tracks the student-teacher role assignment instead of the trivial input-similarity axis.
Adding the projection head makes RINO nearly scale-invariant, with all cross-scale cosines $\geq 0.88$ in the head output, including the unclustered view (Fig.~\ref{fig:scale-similarity}, bottom right).

\begin{figure}[h]
  \centering
  \includegraphics[width=\linewidth]{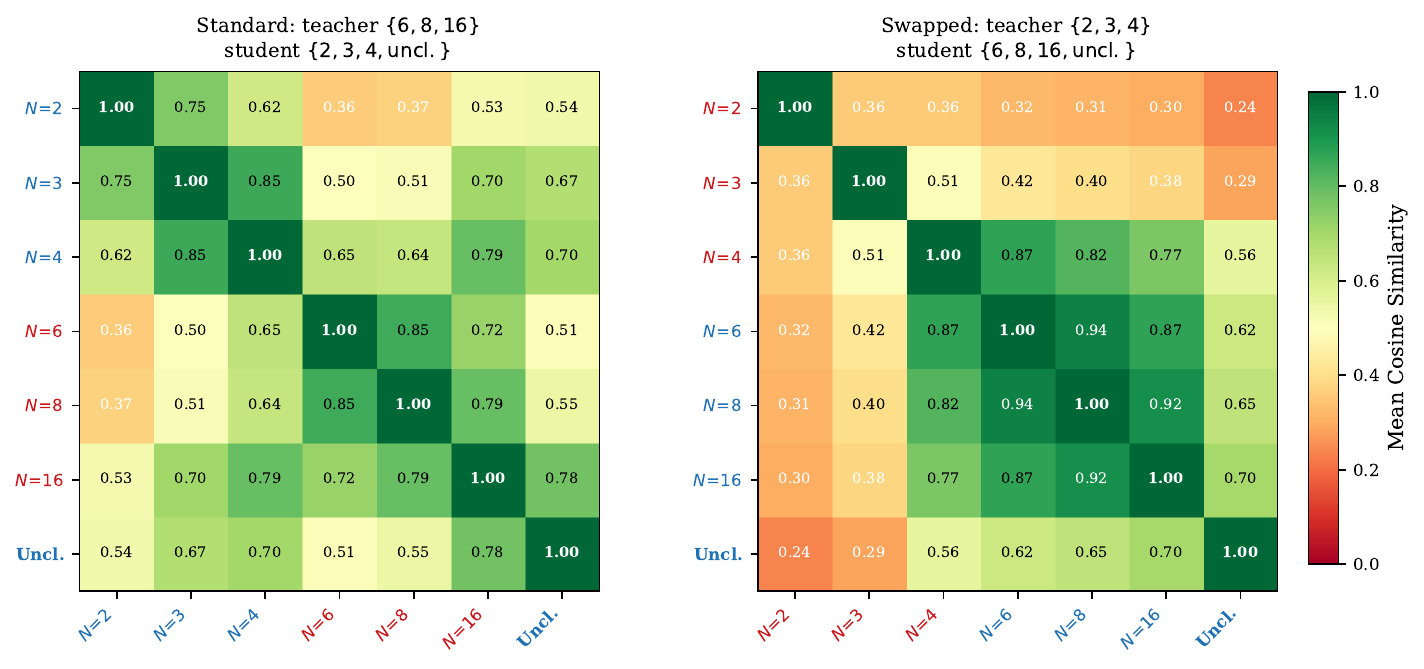}
  \caption{\clstoken{} cosine similarity across \kt{} scales for the standard (left) and swapped (right) teacher-student assignments.
      {\color[HTML]{cb181d}Red} = teacher scales, {\color[HTML]{2171b5}blue} = student scales.
    The grouping of $N\!=\!4$ changes with its role: as a student-only scale (left) it is closest to $N\!=\!3$ (cosine $0.85$, vs.\ $0.65$ to $N\!=\!6$); as a teacher scale (right) it is closest to $N\!=\!6$ (cosine $0.87$, vs.\ $0.51$ to $N\!=\!3$).}
  \label{fig:swapped-cosine}
\end{figure}

\paragraph{Swapped teacher-student roles.}
\label{appendix:swapped-cosine}
To test whether this non-monotonic backbone geometry is causally shaped by the teacher-student role assignment or is a spurious artifact of representation quality, we train a variant with swapped roles: teacher $\{2, 3, 4\}$, with the student additionally seeing $\{6, 8, 16, \mathrm{uncl.}\}$.
Figure~\ref{fig:swapped-cosine} compares the \clstoken{} cosine similarity matrices side by side.

Under the standard assignment, the student-only scale $N\!=\!4$ is closest to $N\!=\!3$ ($0.85$), with cosines $0.65$, $0.64$, and $0.79$ to the teacher scales $N\!=\!6$, $8$, and $16$.
Under the swapped assignment, where $N\!=\!4$ is a teacher scale, it is closest to $N\!=\!6$, $8$, and $16$ ($0.87$, $0.82$, $0.77$) and separates from $N\!=\!3$ ($0.51$). The scales $N\!=\!6$, $8$, $16$ remain mutually similar under both assignments ($0.72$--$0.85$ standard, $0.87$--$0.94$ swapped).
The grouping of $N\!=\!4$ thus follows its role, showing that the backbone geometry is shaped by the role assignment.

\subsection{Cross-Topology Transfer with MLP Head}
\label{appendix:cross-topology-mlp}

Table~\ref{tab:cross-topology-mlp} reports the MLP head variant of the cross-topology evaluation, including the finetuning topology $t \to bqq$.
RINO leads on $t \to bqq$.
On $H \to 4q$, RINO and JetCLR-scale are statistically tied ($0.828 \pm 0.006$ vs.\ $0.824 \pm 0.020$), as are RINO and MPMv2 on $H \to b\bar{b}$ ($0.697 \pm 0.009$ vs.\ $0.692 \pm 0.029$).
JetCLR-scale shows a large gap between MLP and linear heads ($H \to 4q$: $0.824$ vs.\ $0.630$), while RINO's are nearly identical ($0.828$ vs.\ $0.821$), indicating that self-distillation produces linearly accessible class boundaries.

\begin{table}[h]
  \centering
  \caption{Cross-topology transfer ACC (\jetnet{}$\to$\jetclass{}, MLP head).}
  \label{tab:cross-topology-mlp}
  \small
  \begin{tabular}{lccc}
    \toprule
    \textbf{Method}  & $t \to bqq$                & $H \to 4q$                 & $H \to b\bar{b}$           \\
    \midrule
    Supervised       & $0.786 \pm 0.023$          & $0.699 \pm 0.038$          & $0.623 \pm 0.028$          \\
    \midrule
    \multicolumn{4}{l}{\textit{SSL baselines}}                                                              \\
    JetCLR           & $0.700 \pm 0.031$          & $0.696 \pm 0.028$          & $0.686 \pm 0.015$          \\
    MPMv1            & $0.800 \pm 0.019$          & $0.769 \pm 0.006$          & $0.674 \pm 0.017$          \\
    MPMv2            & $0.760 \pm 0.016$          & $0.780 \pm 0.025$          & $0.692 \pm 0.029$          \\
    OmniJet-$\alpha$ & $0.751 \pm 0.050$          & $0.684 \pm 0.028$          & $0.600 \pm 0.017$          \\
    \midrule
    \multicolumn{4}{l}{\textit{\kt{}-view SSL (ours)}}                                                      \\
    JetCLR-scale     & $0.780 \pm 0.023$          & $0.824 \pm 0.020$          & $0.639 \pm 0.014$          \\
    \textbf{RINO}    & $\mathbf{0.847 \pm 0.003}$ & $\mathbf{0.828 \pm 0.006}$ & $\mathbf{0.697 \pm 0.009}$ \\
    \bottomrule
  \end{tabular}
\end{table}

\subsection{\toptagging{}\texorpdfstring{$\to$}{ to }\jetclass{} Transfer Results}
\label{appendix:toptagging}

Table~\ref{tab:tt-results} reports top-tagging accuracy on the \jetclass{} test set for models finetuned on the \toptagging{} dataset~\citep{Kasieczka:2019dbj,TopTaggingDataset}.
This is a transfer between datasets with a smaller domain gap than \jetnet{}$\to$\jetclass{}, not the in-distribution \toptagging{} benchmark, so the values are not comparable to results on the standard \toptagging{} test set.
The leading methods converge to $\sim\!0.93$ accuracy, indicating that this benchmark is less discriminating than the \jetnet{}$\to$\jetclass{} setting used for the main comparison.
Nevertheless, scale factors (App.~\ref{appendix:sf-proxy}), which directly quantify how strongly a classifier depends on simulation-specific structure, separate the methods, with RINO's closest to unity.

\begin{table}[h]
  \centering
  \caption{\toptagging{}$\to$\jetclass{} transfer: top-tagging accuracy ($\mu \pm \sigma$ over $6$ seeds) on the \jetclass{} test set of models finetuned on \toptagging{}. These are transfer results, not comparable to in-distribution results on the standard \toptagging{} test set. The leading methods saturate near $0.93$. Bold marks the best per column; \underline{underline} marks entries within $1\sigma$ of the best.}
  \label{tab:tt-results}
  \small
  \begin{tabular}{lcc}
    \toprule
    \textbf{Method}                       & \textbf{MLP head}                   & \textbf{Linear head}                \\
    \midrule
    Supervised                            & $0.929 \pm 0.001$                   & $0.929 \pm 0.001$                   \\
    \midrule
    \multicolumn{3}{l}{\textit{SSL baselines}}                                                                        \\
    JetCLR~\citep{JetCLR}                  & $0.921 \pm 0.001$                   & $0.921 \pm 0.001$                   \\
    MPMv1~\citep{MPMv1}                    & $0.929 \pm 0.002$                   & $0.927 \pm 0.001$                   \\
    MPMv2~\citep{MPMv2}                    & $0.932 \pm 0.001$                   & $\underline{0.933 \pm 0.001}$       \\
    OmniJet-$\alpha$~\citep{omnijet-alpha} & $0.760 \pm 0.009$                   & $0.756 \pm 0.008$                   \\
    \midrule
    \multicolumn{3}{l}{\textit{\kt{}-view SSL (ours)}}                                                                \\
    JetCLR-scale                          & $\mathbf{0.936 \pm 0.003}$          & $\mathbf{0.935 \pm 0.003}$          \\
    \textbf{RINO}                         & $\underline{0.933 \pm 0.001}$       & $\underline{0.933 \pm 0.001}$       \\
    \bottomrule
  \end{tabular}
\end{table}

Table~\ref{tab:tt-cross-topology} extends this to cross-topology transfer: models finetuned on \toptagging{} for top tagging are evaluated on the unseen $H \to 4q$ signal from \jetclass{}.
When finetuned on \toptagging{} (smaller domain gap with overlapping $p_\mathrm{T}$ range), cross-topology transfer is much more uniform: the four methods evaluated span $H \to 4q$ AUC $0.877$--$0.900$, narrower than the $0.803$--$0.894$ of the same four methods on \jetnet{}$\to$\jetclass{} (Table~\ref{tab:cross-topology}), and RINO ($0.896 \pm 0.002$) is within one standard deviation of the leading MPMv2 ($0.900 \pm 0.005$).
This contrasts with \jetnet{}$\to$\jetclass{} (Table~\ref{tab:cross-topology}), where RINO leads by $+4.1\%$, consistent with its advantage being specific to larger domain shifts.

\begin{table}[h]
  \centering
  \caption{\toptagging{}$\to$\jetclass{} cross-topology AUC (linear head, 6 seeds).
    With the smaller domain gap, the four methods reach similar AUC.}
  \label{tab:tt-cross-topology}
  \small
  \begin{tabular}{lcc}
    \toprule
    \textbf{Method} & $t \to bqq$                & $H \to 4q$                 \\
    \midrule
    MPMv2           & $\mathbf{0.982 \pm 0.001}$ & $\mathbf{0.900 \pm 0.005}$ \\
    RINO            & $0.981 \pm 0.000$          & $0.896 \pm 0.002$          \\
    MPMv1           & $0.980 \pm 0.000$          & $0.889 \pm 0.001$          \\
    JetCLR          & $0.976 \pm 0.001$          & $0.877 \pm 0.005$          \\
    \bottomrule
  \end{tabular}
\end{table}

\subsection{Scale-Factor Proxy (\toptagging{}\texorpdfstring{$\to$}{ to }\jetclass{})}
\label{appendix:sf-proxy}

Scale factors (SFs) are a standard downstream component of collider analyses: efficiencies measured in data control regions are divided by their simulated values and used to correct simulated yields (App.~\ref{appendix:deployment}).
They therefore directly quantify how strongly a classifier depends on simulation-specific structure.
We additionally use the \toptagging{} dataset to construct a proxy experiment that measures them, using \toptagging{} as source (``simulation'') and \jetclass{} as target (``data''), with the \toptagging{}-finetuned models of Table~\ref{tab:tt-results}.
We partition the source signal-score cumulative distribution function (CDF) into six disjoint intervals (the last containing the highest-scoring $5\%$ of jets), apply the same score boundaries to the target, and compute the target/source efficiency ratio in each.
For each finetuning run, let $s$ be the signal score, with CDF $F$ over the \toptagging{} test signal jets and $w$-quantile $q_w$:
\begin{equation}
  F(s) = P_\mathrm{src}\!\left(s' \le s \mid \text{signal}\right),
  \qquad
  q_w = F^{-1}(w),
  \label{eq:sf-cdf}
\end{equation}
with $q_0=-\infty$ and $q_1=+\infty$.
An interval $[a,b)$ thus selects jets with $q_a \le s < q_b$, \ie source signal efficiencies $1 - F(s)$ between $1-b$ and $1-a$; the last interval, $[0.95, 1]$, is the tightest working point ($1 - F(s) \le 0.05$).
For each interval $[a,b)$, the signal efficiency in domain $d \in \{\mathrm{src}, \mathrm{tgt}\}$ and the scale factor are
\begin{equation}
  \varepsilon^{d}_{[a,b)} = P_{d}\!\left(q_a \le s < q_b \mid \text{signal}\right),
  \qquad
  \mathrm{SF}_{[a,b)} = \frac{\varepsilon^{\mathrm{tgt}}_{[a,b)}}{\varepsilon^{\mathrm{src}}_{[a,b)}},
  \label{eq:sf}
\end{equation}
so $\mathrm{SF}=1$ means that the target populates that interval exactly as the source does.
Table~\ref{tab:sf-proxy} lists the scale factors per bin and their mean absolute departure from unity.

\begin{table}[h]
  \centering
  \caption{\toptagging{}$\to$\jetclass{} scale factors (Eq.~\ref{eq:sf}) per interval of the source signal-score CDF (the last containing the highest-scoring $5\%$ of jets) and their mean absolute departure from unity (last column). Entries are mean $\pm$ standard deviation over finetuning runs; bold marks the value closest to unity per bin.}
  \label{tab:sf-proxy}
  \small
  \setlength{\tabcolsep}{4pt}
  \resizebox{\linewidth}{!}{%
  \begin{tabular}{lccccccc}
    \toprule
    \textbf{Method}  & $[0, 0.3)$                  & $[0.3, 0.5)$                & $[0.5, 0.7)$                & $[0.7, 0.9)$                & $[0.9, 0.95)$               & $[0.95, 1]$                 & $\langle|\mathrm{SF}-1|\rangle \downarrow$ \\
    \midrule
    Supervised       & $1.120 \pm 0.053$           & $0.947 \pm 0.022$           & $0.933 \pm 0.023$           & $0.960 \pm 0.041$           & $0.958 \pm 0.073$           & $\mathbf{0.972 \pm 0.128}$  & $0.059$                                    \\
    \midrule
    \multicolumn{8}{l}{\textit{SSL baselines}}                                                                                                                                                                                                   \\
    JetCLR           & $1.134 \pm 0.032$           & $0.942 \pm 0.012$           & $0.903 \pm 0.020$           & $0.898 \pm 0.023$           & $0.964 \pm 0.029$           & $1.256 \pm 0.084$           & $0.114$                                    \\
    MPMv1            & $1.162 \pm 0.047$           & $\mathbf{0.966 \pm 0.013}$  & $0.892 \pm 0.014$           & $0.884 \pm 0.030$           & $0.947 \pm 0.043$           & $1.117 \pm 0.083$           & $0.098$                                    \\
    MPMv2            & $0.928 \pm 0.024$           & $0.867 \pm 0.021$           & $0.898 \pm 0.008$           & $1.019 \pm 0.022$           & $1.278 \pm 0.048$           & $1.999 \pm 0.112$           & $0.267$                                    \\
    OmniJet-$\alpha$ & $1.073 \pm 0.029$           & $0.938 \pm 0.020$           & $0.949 \pm 0.031$           & $0.975 \pm 0.026$           & $\mathbf{1.001 \pm 0.046}$  & $1.110 \pm 0.159$           & $0.054$                                    \\
    \midrule
    \multicolumn{8}{l}{\textit{\kt{}-view SSL (ours)}}                                                                                                                                                                                           \\
    JetCLR-scale     & $\mathbf{0.980 \pm 0.051}$  & $0.871 \pm 0.030$           & $0.884 \pm 0.020$           & $0.980 \pm 0.028$           & $1.186 \pm 0.086$           & $1.967 \pm 0.282$           & $0.240$                                    \\
    \textbf{RINO}    & $1.029 \pm 0.057$           & $0.963 \pm 0.027$           & $\mathbf{0.978 \pm 0.018}$  & $\mathbf{1.001 \pm 0.059}$  & $1.031 \pm 0.084$           & $1.029 \pm 0.115$           & $\mathbf{0.025}$                           \\
    \bottomrule
  \end{tabular}%
  }
\end{table}

RINO's $2.5\%$ average deformation is less than half the next-lowest value ($5.4\%$), indicating substantially less dependence on simulation-specific structure.

\subsection{Path to Experimental Data}
\label{appendix:deployment}
\begin{figure}[h]
  \centering
  \begin{tikzpicture}[
      stage/.style={draw, rounded corners=2pt, align=center, text width=0.27\linewidth, minimum height=1.75cm, inner sep=3pt, font=\small},
      proxy/.style={align=center, text width=0.27\linewidth, inner sep=1pt, font=\footnotesize, text=black!70},
      arr/.style={-{Stealth[length=2mm]}, thick}
    ]
    \node[stage, fill=blue!6] (pre) {\textbf{1. Pretrain on}\\ \textbf{collision data}\\ unlabeled jets};
    \node[stage, fill=orange!8, right=0.45cm of pre] (ft) {\textbf{2. Finetune on}\\ \textbf{labeled simulations}\\ signal and background};
    \node[stage, fill=green!8, right=0.45cm of ft] (cal) {\textbf{3. Calibrate in data}\\ scale factors and\\ uncertainties from\\ control regions};
    \draw[arr] (pre) -- (ft);
    \draw[arr] (ft) -- (cal);
    \node[proxy, below=0.12cm of pre] {This work: \jetclass{} QCD (simulated)};
    \node[proxy, below=0.12cm of ft] {This work: \jetnet{}, \toptagging{}};
    \node[proxy, below=0.12cm of cal] {This work: labeled top and QCD test jets from \mbox{\jetclass{}}; SF proxy};
  \end{tikzpicture}
  \caption{Intended deployment of RINO. Labels enter only in stage~2, and stage~3 is the standard calibration used for supervised taggers. The bottom row lists the simulated stand-ins used in this work, where \jetclass{} top jets and all \jetclass{} labels enter only in stage~3, for evaluation.}
  \label{fig:deployment}
\end{figure}
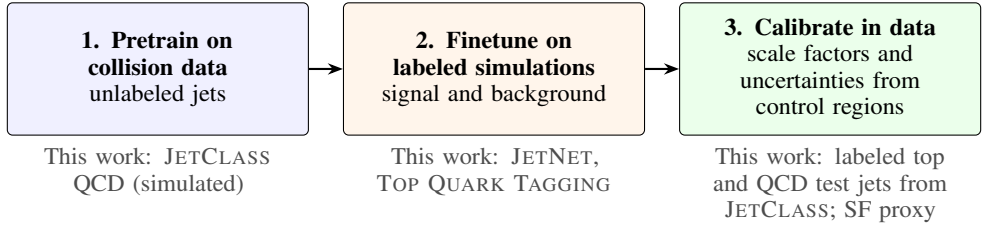

Figure~\ref{fig:deployment} shows the intended use of RINO in an experimental analysis, in three stages.

\paragraph{1. Pretrain on collision data.}
RINO can learn jet representations from unlabeled experimental data, since its pretraining uses no labels. Nevertheless, in this work, for more straightforward evaluations, pretraining uses simulated \jetclass{} QCD jets to mimic the QCD-dominated data. 
Including a fraction of simulated jets and also specialized control regions to isolate more signal-like topologies for the pretraining are both interesting areas for future work.

\paragraph{2. Finetune on labeled simulations.}
The pretrained backbone is finetuned on simulated signal and background jets, where labels are available, as in our \jetnet{} and \toptagging{} experiments.

\paragraph{3. Calibrate in data.}
Signal efficiencies~\citep{CMS-DP-2025-010} and background mistag rates are measured in dedicated control regions.
For top tagging, semileptonic-\ttbar{} tag-and-probe can measure signal efficiency, while QCD-enriched multijet regions constrain mistag rates.
The resulting data-to-simulation scale factors (SFs) correct the predicted yields; detector, shower, and hadronization variations enter the final likelihood as systematic uncertainties.
Appendix~\ref{appendix:sf-proxy} gives example SFs measured in a simulation-to-simulation proxy.

\subsection{Full Data-Efficiency Curves}
\label{app:data-efficiency}

\begin{figure}[h]
  \centering
  \includegraphics[width=0.85\linewidth]{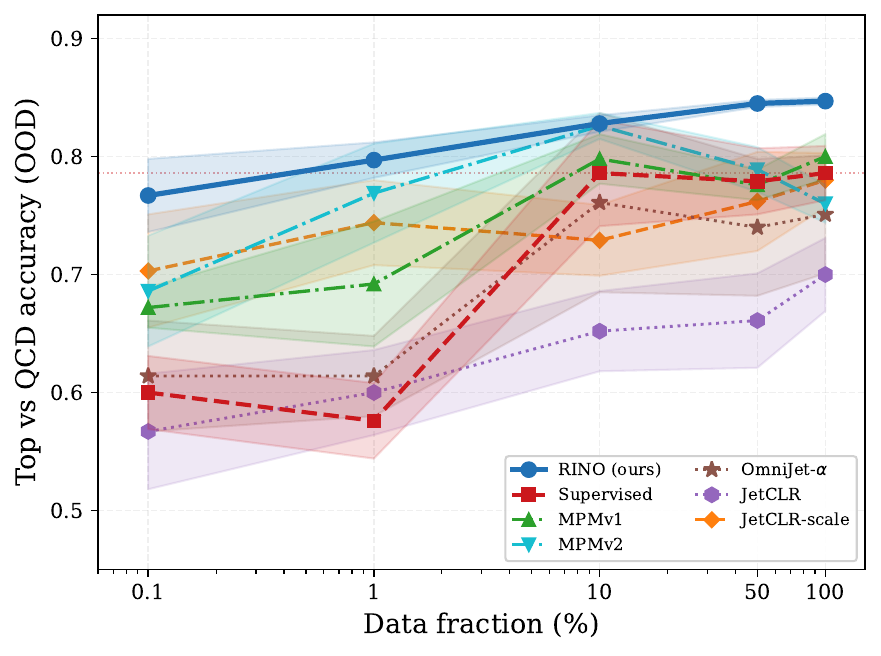}
  \caption{Full data-efficiency curves (all seven methods) with $\pm 1\sigma$ bands over $10$ seeds.}
  \label{fig:le-full}
\end{figure}

Figure~\ref{fig:le-full} shows the complete data-efficiency curves for all seven methods evaluated in Sec.~\ref{sec:exp:data-efficiency}.
This includes the JetCLR-scale variant, which isolates the effect of replacing JetCLR's generic augmentations with \kt{} scale views within the same contrastive objective family.
OmniJet-$\alpha$ is omitted from the main-text figure for clarity; its behavior is qualitatively similar to most other baselines (non-monotonic, high variance).

\paragraph{Non-monotonic behavior under distribution shift.}
Several baselines exhibit non-monotonic data-efficiency curves: MPMv2 peaks at $10\%$ and declines at $50$--$100\%$. The mechanism is discussed in Sec.~\ref{sec:exp:data-efficiency}.
To rule out a batch-size artifact, we re-ran MPMv2 at $50\%$ with batch size $256$ (vs.\ the default $1024$): accuracy changed from $0.789$ to $0.761 \pm 0.027$, indicating that the non-monotonicity is not simply a batch-size effect.
We would like to emphasize that the supervised baseline gains no \jetclass{} accuracy beyond $10\%$ of the \jetnet{} labels ($0.786$, $0.779$, and $0.786$ at $10\%$, $50\%$, and $100\%$), which is not classical overfitting: its held-out in-distribution \jetnet{} (source) accuracy stays flat or rises with the label fraction.
The model is thus not memorizing its training set but specializing to simulation-specific features that do not transfer well.
RINO is designed to prevent this: physically, \kt{} coarsening absorbs low-scale radiation into harder subjets~\citep{Salam:2010nqg}, so requiring agreement across scales discourages reliance on soft, low-scale details, precisely where parton-shower and hadronization modeling is approximate and generator-dependent~\citep{Nagy:2017ggp,Barnard:2016qma}.
RINO improves monotonically across all label fractions and leads at every fraction; the only other monotonic curve, JetCLR, stays far lower ($0.700 \pm 0.031$ at $100\%$). This is consistent with RINO's ${\sim}3\times$ smaller domain shift in representation space (Table~\ref{tab:domain-shift}).

\subsection{MLP vs.\ Linear Head Diagnostic}
\label{appendix:mlp-linear}

Table~\ref{tab:mlp-linear} compares MLP and linear head performance on the \jetnet{}$\to$\jetclass{} OOD benchmark.
The significance column reports the Welch $t$-statistic $t = (\mu_\mathrm{MLP} - \mu_\mathrm{Linear}) / \sqrt{\sigma_\mathrm{MLP}^2/n + \sigma_\mathrm{Linear}^2/n}$ with $n=10$ seeds per condition. Degrees of freedom follow the Welch--Satterthwaite approximation ($\approx 11$ for RINO, whose two variances differ, and $\approx 12$--$17$ for the other methods).
RINO is the only method with a significantly positive $t$ ($+3.0$, two-tailed $p\approx 0.01$), indicating exploitable nonlinear structure; JetCLR and MPMv2 show clear degradation ($t < -3$, $p<10^{-3}$).

\begin{table}[h]
  \centering
  \caption{MLP vs.\ linear head on OOD top tagging (\jetnet{}$\to$\jetclass{}).}
  \label{tab:mlp-linear}
  \small
  \begin{tabular}{lccc}
    \toprule
    \textbf{Method}  & \textbf{MLP}               & \textbf{Linear}            & $t$             \\
    \midrule
    \textbf{RINO}    & $\mathbf{0.847 \pm 0.003}$ & $\mathbf{0.838 \pm 0.009}$ & $\mathbf{+3.0}$ \\
    Supervised       & $0.786 \pm 0.023$          & $0.784 \pm 0.033$          & $+0.2$          \\
    MPMv1            & $0.800 \pm 0.019$          & $0.809 \pm 0.008$          & $-1.4$          \\
    MPMv2            & $0.760 \pm 0.016$          & $0.817 \pm 0.027$          & $-5.7$          \\
    JetCLR           & $0.700 \pm 0.031$          & $0.791 \pm 0.048$          & $-5.0$          \\
    JetCLR-scale     & $0.780 \pm 0.023$          & $0.803 \pm 0.017$          & $-2.5$          \\
    OmniJet-$\alpha$ & $0.751 \pm 0.050$          & $0.749 \pm 0.080$          & $+0.1$          \\
    \bottomrule
  \end{tabular}
\end{table}

\subsection{Domain Shift in Representation Space}
\label{appendix:domain-shift}

\paragraph{Protocol.} Both metrics are computed on $5{,}000$ jets per domain. $W_1$ is averaged over embedding dimensions; the squared maximum mean discrepancy (MMD$^2$, biased estimator) uses an RBF kernel $\exp(-\gamma\lVert x - y\rVert^2)$ with fixed $\gamma = 1/d$, where $d$ is the embedding dimension.

Table~\ref{tab:domain-shift-pretrained} reports the same metrics as Table~\ref{tab:domain-shift} on the pretrained backbones, before finetuning, with signed changes indicating whether finetuning increases ($+$) or decreases ($-$) the gap.
RINO already has the smallest pretrained shift on three of the four metrics (\eg $W_1$ (QCD): $0.055$, vs.\ at least $0.092$ for the baselines).
JetCLR-scale, which shares RINO's \kt{} views, is the other method with a small pretrained shift: the two \kt{}-view methods have the two smallest values on \emph{every} metric.
JetCLR-scale's shift increases during finetuning but stays relatively low (Table~\ref{tab:domain-shift}), while RINO's decreases further on all four metrics (\eg $W_1$ (QCD): $0.055 \to 0.043$).
MPMv2 is the only other method whose gap decreases on all four.
The remaining baselines do not consistently benefit from finetuning: the changes are mixed for JetCLR and MPMv1.

These trends connect to the feature-distortion picture of \citet{LP-FT} (Sec.~\ref{sec:exp:data-efficiency}): finetuning updates the features of in-distribution (\jetnet{}) jets more than those of OOD (\jetclass{}) jets, which can widen the gap between the two domains.
For RINO, whose pretrained representations are already close across domains, finetuning instead brings them even closer together.
This suggests that the small pretrained shift comes from the \kt{} views, and provides further evidence for the physical picture of App.~\ref{app:data-efficiency}: \kt{} coarsening absorbs low-scale radiation into harder subjets~\citep{Salam:2010nqg}, so requiring agreement across scales discourages reliance on soft, low-scale details, where parton-shower and hadronization modeling is approximate and generator-dependent~\citep{Nagy:2017ggp,Barnard:2016qma}.

\begin{table}[h]
  \centering
  \caption{Domain shift in \emph{pretrained} representations (\jetnet{} vs.\ \jetclass{}), before finetuning. Colored values give the change \textit{after} finetuning (Table~\ref{tab:domain-shift}): $-$ (green) = finetuning reduces the gap, $+$ (red) = finetuning increases it.}
  \label{tab:domain-shift-pretrained}
  \setlength{\tabcolsep}{4pt}
  \begin{tabular}{lcccc}
    \toprule
    \textbf{Method}  & $W_1$ (QCD) $\downarrow$ & $W_1$ (top) $\downarrow$ & MMD$^2$ (QCD) $\downarrow$ & MMD$^2$ (top) $\downarrow$ \\
    \midrule
    JetCLR           & $0.128$ {\scriptsize\color{red}$+.004$} & $0.125$ {\scriptsize\color{red}$+.002$} & $0.223$ {\scriptsize\color{ForestGreen}$-.027$} & $0.212$ {\scriptsize\color{ForestGreen}$-.021$} \\
    OmniJet-$\alpha$ & $0.181$ {\scriptsize\color{red}$+.634$} & $0.246$ {\scriptsize\color{red}$+.506$} & $0.095$ {\scriptsize\color{red}$+.476$} & $0.193$ {\scriptsize\color{red}$+.359$} \\
    MPMv1            & $0.232$ {\scriptsize\color{red}$+.009$} & $0.274$ {\scriptsize\color{ForestGreen}$-.012$} & $0.120$ {\scriptsize\color{red}$+.013$} & $0.183$ {\scriptsize\color{ForestGreen}$-.021$} \\
    MPMv2            & $0.409$ {\scriptsize\color{ForestGreen}$-.034$} & $0.374$ {\scriptsize\color{ForestGreen}$-.026$} & $0.427$ {\scriptsize\color{ForestGreen}$-.089$} & $0.409$ {\scriptsize\color{ForestGreen}$-.104$} \\
    \midrule
    JetCLR-scale     & $0.092$ {\scriptsize\color{red}$+.041$} & $0.071$ {\scriptsize\color{red}$+.074$} & $0.017$ {\scriptsize\color{red}$+.028$} & $\mathbf{0.009}$ {\scriptsize\color{red}$+.039$} \\
    \textbf{RINO}    & $\mathbf{0.055}$ {\scriptsize\color{ForestGreen}$-.012$} & $\mathbf{0.052}$ {\scriptsize\color{ForestGreen}$-.008$} & $\mathbf{0.012}$ {\scriptsize\color{ForestGreen}$-.005$} & $0.011$ {\scriptsize\color{ForestGreen}$-.004$} \\
    \bottomrule
  \end{tabular}
\end{table}

\subsection{Baseline Implementation Details}
\label{appendix:baselines}

All SSL baselines are implemented under a controlled implementation and finetuning setup so that the main variables across methods are the pretraining objective and view-construction strategy.
This setup reduces confounders from depth, width, normalization, and activation choices that would otherwise conflate model capacity with pretraining quality.
Table~\ref{tab:baseline-details} summarizes the key implementation choices.
We do not include J-JEPA~\citep{J-JEPA} or HEP-JEPA~\citep{HEP-JEPA} in our main comparison: J-JEPA reports modest gains over training from scratch, and HEP-JEPA's code has not been released, precluding direct reproduction in our controlled-backbone protocol.
Among existing methods, J-JEPA is the closest conceptual relative of RINO: it reclusters each jet into C/A subjets and predicts the embeddings of target subjets from context subjets within a single jet, but does not align representations \emph{across} different coarse-graining scales.
Our ablation (Sec.~\ref{sec:exp:ablation}) shows that cross-scale \kt{} views account for the majority of the gain ($+11.4\%$) regardless of the SSL family, suggesting that the multi-scale view construction dominates the choice of joint-embedding objective.

\begin{table}[h]
  \centering
  \caption{Baseline implementation summary. All methods share the same 7 kinematic input features (Table~\ref{tab:particle-features}), the same QCD pretraining data, and the same finetuning recipe (App.~\ref{appendix:finetune-protocol}).}
  \label{tab:baseline-details}
  \small
  \setlength{\tabcolsep}{3.5pt}
  \begin{tabular}{@{}lllll@{}}
    \toprule
    \textbf{Method}  & \textbf{Backbone} & \textbf{Attention} & \textbf{Input} & \textbf{Loss}    \\
    \midrule
    RINO             & JTE enc.          & Bidir.             & Continuous     & DINO+iBOT        \\
    Supervised       & JTE enc.          & Bidir.             & Continuous     & Cross-ent.       \\
    MPMv1            & JTE enc.          & Bidir.             & Continuous     & Masked VQ pred.  \\
    MPMv2            & JTE enc.          & Bidir.             & Continuous     & Masked K-means + L1 + flow \\
    JetCLR           & JTE enc.          & Bidir.             & Continuous     & NT-Xent          \\
    JetCLR-scale     & JTE enc.          & Bidir.             & Continuous     & NT-Xent+MAE      \\
    OmniJet-$\alpha$ & JTD dec.          & Causal             & VQ tokens      & Next-token pred. \\
    \bottomrule
  \end{tabular}
\end{table}

\paragraph{Shared components.}
The jet transformer encoder (JTE) backbone is identical across all encoder-based methods: $8$ layers, $d_\mathrm{model}\!=\!256$, $16$ heads, SwiGLU FFN, RMSNorm, LayerScale, and $4$ register tokens.
MPMv1 and OmniJet-$\alpha$ share a VQ-VAE tokenizer with a $512$-code codebook, trained once on the QCD pretraining set and frozen during SSL pretraining.
All methods are pretrained on the same QCD-only data.

\paragraph{Hyperparameter parity.}
Each baseline uses the core hyperparameters from its original publication (loss weights, masking ratios, temperature), which we assume are near-optimal for the respective objectives.
RINO's teacher temperature and EMA momentum were selected on the OOD top-vs-QCD accuracy on \jetclass{}.
This step contributes $+1.0\%$ (Table~\ref{tab:ablation}), and RINO before tuning ($0.839 \pm 0.005$) already exceeds every baseline's MLP-head accuracy in Table~\ref{tab:main-comparison} (best: MPMv1, $0.800 \pm 0.019$).
All encoder-based methods share the same backbone, and all methods share the same data and finetuning protocol, so the comparison isolates the pretraining objective and view construction as much as possible.

\paragraph{OmniJet-$\alpha$.}
OmniJet-$\alpha$ is the only decoder-based method.
The original model~\citep{omnijet-alpha} uses a $3$-layer GPT-style decoder with $8$ heads and an $8{,}194$-token vocabulary.
Our adaptation replaces this with an $8$-layer jet transformer decoder (JTD) that mirrors the JTE in depth, width, normalization, and activation, differing only in the causal attention mask required by the autoregressive objective.
This controls for model capacity when comparing the decoder baseline to encoder-based methods.
We use the shared $512$-code VQ-VAE tokenizer rather than the original vocabulary.
Because causal attention imposes an ordering on an inherently unordered set (jet constituents have no natural sequence), the decoder can only aggregate information left-to-right.
We use mean pooling over all token outputs as the jet representation, which empirically outperforms the \texttt{last\_token} alternative.

\paragraph{MPMv1 and MPMv2.}
Both follow the masked-particle modeling framework of \citet{MPMv1} and \citet{MPMv2}, adapted to use the shared JTE backbone.
The backbone receives \emph{continuous} input features (the same 7 kinematic features as RINO), projected through a linear embedding layer.
MPMv1 uses the VQ-VAE only as a \emph{target generator} for a cross-entropy classification loss at masked positions.
MPMv2 uses no VQ-VAE: at masked positions, it predicts each constituent's cluster in the $16{,}384$-centroid K-means codebook released with \citet{MPMv2}, regresses the continuous features directly (L1 loss), and models them with a conditional normalizing flow.
Mean pooling is used because MPM's per-particle masked-prediction objective provides no training signal for a \clstoken{}.

\paragraph{JetCLR and JetCLR-scale.}
JetCLR~\citep{JetCLR} uses the NT-Xent contrastive loss with standard jet augmentations (rotations, translations, $p_\mathrm{T}$-dependent smearing, and collinear splittings).
JetCLR-scale isolates the contribution of multi-scale view construction: it replaces JetCLR's generic augmentations with the same \kt{} clustering views used by RINO, while keeping a contrastive objective.
Scales are grouped into three tiers: coarse ($N \in \{2,3,4\}$), intermediate ($N \in \{6,8\}$), and fine ($N = 16$ or unclustered). Each training batch samples one scale from each of two randomly chosen tiers, producing a two-view pair with a guaranteed cross-scale gap (\eg $N\!=\!3$ vs.\ $N\!=\!16$).
The NT-Xent loss encourages agreement across these paired scales, and an MAE reconstruction auxiliary at the fine scale stabilizes low-data training.
The JetCLR-to-JetCLR-scale comparison ($0.700$ vs.\ $0.780$) confirms that the \kt{} views account for most of the gain.

\subsection{Compute}
\label{appendix:compute}
Pretraining a single RINO model on the QCD subset of \jetclass{} (${\sim}10^7$ jets, $100$ epochs) takes approximately $45$ minutes per epoch on $2\times$ NVIDIA A100-80GB, for a wall-clock cost of ${\sim}80$ hours per pretraining run.
Finetuning on \jetnet{} takes ${\sim}2$ minutes per epoch on a single NVIDIA 3090 or A10, for at most $55$ epochs with early stopping; runs typically converge within ${\sim}25$ epochs (${\sim}50$ minutes wall-clock per seed).
Finetuning on the \toptagging{} dataset takes ${\sim}15$ minutes per epoch on $2\times$ 3090/A10, again capped at $55$ epochs with early stopping (${\sim}25$ epochs and ${\sim}6$ wall-clock hours per seed in practice).

Table~\ref{tab:pretrain-cost} compares per-method pretraining wall-clock cost on the same QCD-only \jetclass{} subset.
All encoder-based methods use the shared JTE backbone (OmniJet-$\alpha$ uses the matched JTD).
MPMv1 and OmniJet-$\alpha$ additionally require a VQ-VAE tokenizer (${\sim}8$h, trained once and shared).

\begin{table}[h]
\centering
\caption{Per-method pretraining wall-clock cost (QCD-only \jetclass{} subset, estimated for $100$ epochs).
$\dagger$~Requires a shared VQ-VAE tokenizer (${\sim}8$h additional, trained once).}
\label{tab:pretrain-cost}
\small
\begin{tabular}{lcc}
\toprule
\textbf{Method} & \textbf{Time/epoch (min)} & \textbf{Total (h)} \\
\midrule
JetCLR              & ${\sim}10$ & ${\sim}20$  \\
JetCLR-scale        & ${\sim}25$ & ${\sim}45$  \\
MPMv1$^\dagger$     & ${\sim}30$ & ${\sim}50$  \\
MPMv2               & ${\sim}40$ & ${\sim}65$  \\
OmniJet-$\alpha$$^\dagger$ & ${\sim}20$ & ${\sim}30$ \\
iBOT-only           & ${\sim}50$ & ${\sim}80$  \\
\textbf{RINO}       & ${\sim}45$ & ${\sim}80$  \\
\bottomrule
\end{tabular}
\end{table}

\end{document}